\documentclass[11pt]{article}
\pdfoutput=1
\usepackage[utf8]{inputenc}
\UseRawInputEncoding
\usepackage{rotating}
\usepackage{verbatim,amsmath,amssymb}
\usepackage{epsfig,float,color}
\usepackage{geometry}
\usepackage{setspace}
\usepackage[numbers,sort&compress]{natbib}
\usepackage{gensymb}
\usepackage{xcolor}
\usepackage{bbding}
\usepackage{amsthm,mathrsfs,amsfonts,dsfont}
\usepackage{hyperref}
\usepackage[capitalise]{cleveref}
\usepackage[yyyymmdd,hhmmss]{datetime}
\usepackage{cleveref}
\usepackage{caption}
\usepackage{subcaption}
\usepackage{tablefootnote}
\usepackage{mmap}
\usepackage[section]{placeins}
\definecolor{mypink}{RGB}{255, 20, 147}
\definecolor{mygreen}{RGB}{34, 139, 34}
\definecolor{myolive}{RGB}{65, 105, 225}
\usepackage[utf8]{inputenc}
\usepackage{threeparttable}
\usepackage{booktabs, caption, makecell}

\definecolor{darkblue}{rgb}{0,0,1}
\hypersetup{pdftex=true, colorlinks=true, breaklinks=true, linkcolor=darkblue, menucolor=darkblue, pagecolor=darkblue, citecolor=darkblue, urlcolor=darkblue}

\newcommand{\bitm}{\begin{itemize}}
\newcommand{\eitm}{\end{itemize}}
\newcommand{\bnumr}{\begin{enumerate}}
\newcommand{\enumr}{\end{enumerate}}

\newcommand {\sigab}{\sigma^{\alpha\beta}}

\newcommand {\auab}{a_{\alpha\beta}}

\newcommand {\buab}{b_{\alpha\beta}}

\newcommand {\tauab}{\tau^{\alpha\beta}}

\newcommand {\eqb}[1]{\begin{equation}\begin{array}{#1}}
\newcommand {\eqe}{\end{array}\end{equation}}

\newcommand {\esb}[1]{\begin{equation*}\begin{array}{#1}}
\newcommand {\ese}{\end{array}\end{equation*}}
\newcommand {\ds}{\displaystyle}

\newcommand {\pa}[2]{\frac{\partial{#1}}{\partial{#2}}}

\newcommand {\back}{\! \! \!}
\newcommand {\is}{\back &=& \back}
\newcommand {\dis}{\back &:=& \back}

\newcommand {\mi}{\back &-& \back}

\newcommand {\dif}{\mathrm{d}}

\newcommand {\II}{{I\kern-.3em I}}
\newcommand {\III}{{I\kern-.3em I\kern-.3em I}}

\newcommand {\mrs}{\mathrm{s}}

\newcommand {\mf}{\mathbf{f}}

\newcommand {\mr}{\mathbf{r}}

\newcommand {\muu}{\mathbf{u}}

\newcommand {\ba}{\boldsymbol{a}}

\newcommand {\bn}{\boldsymbol{n}}

\newcommand {\br}{\boldsymbol{r}}

\newcommand {\bt}{\boldsymbol{t}}

\newcommand {\bx}{\boldsymbol{x}}

\newcommand {\mK}{\mathbf{K}}

\newcommand {\mM}{\mathbf{M}}
\newcommand {\mN}{\mathbf{N}}

\newcommand {\bF}{\boldsymbol{F}}

\newcommand {\bY}{\boldsymbol{Y}}

\newcommand {\sig}{\sigma}

\newcommand {\IR}{{\rm\kern.24em
   \vrule width.02em height1.53ex depth-.05ex
   \kern-.3em R}}
\newcommand {\ic}{{\rm\kern.20em
   \vrule width.02em height1.0ex depth-.05ex
   \kern-.22em c}}
\newcommand {\ia}{{\rm\kern.20em
   \vrule width.02em height1.05ex depth-.0ex
   \kern-.25em a}}
\newcommand {\IC}{{\rm\kern.24em
   \vrule width.02em height1.4ex depth-.05ex
   \kern-.26em C}}
\newcommand {\ID}{{\rm\kern.34em
   \vrule width.02em height1.5ex depth-.05ex
   \kern-.36em D}}
\newcommand {\IS}{{\rm\kern.24em
   \vrule width.02em height1.6ex depth.05ex
   \kern-.26em S}}
\newcommand {\IT}{{\rm\kern.50em
   \vrule width.02em height1.55ex depth-.05ex
   \kern-.52em T}}

\newcommand {\IE}{{\rm\kern.24em
   \vrule width.02em height1.55ex depth-.05ex
   \kern-.33em E}}
\newcommand {\IEa}{{\rm\kern.24em
   \vrule width.02em height1.55ex depth-.05ex
   \kern-.33em E}^{1}_{ijkl}}
\newcommand {\IEb}{{\rm\kern.24em
   \vrule width.02em height1.55ex depth-.05ex
   \kern-.33em E}^{2}_{ijkl}}

\newcommand {\sJ}{\mathcal{J}}

\newcommand {\Ass}[2]{\kern 0.9ex \vrule width0.45em height0.2ex depth0ex \kern -2.1ex \bigwedge_{#1}^{#2}}
\newcommand {\ASS}[2]{\kern 1.45ex \vrule width0.5em height0.2ex depth0ex \kern -2.65ex \bigwedge_{#1}^{#2}}

\graphicspath{ {images/} }

\begin{document}

\begin{center}
\Large{\bf{Comparing quantum, molecular and continuum models for graphene at large deformations }}\\

\end{center}

\begin{center}

\large{Aningi Mokhalingam$^\dag$\footnote{Email: aningi@iitk.ac.in},
       Reza Ghaffari$^\S$\footnote{Email: ghaffari@aices.rwth-aachen.de},
       Roger A. Sauer$^\S$\footnote{Email: sauer@aices.rwth-aachen.de}
       and Shakti S. Gupta$^\dag$\footnote{Corresponding author, email: ssgupta@iitk.ac.in}}\\
\vspace{4mm}

\small{\textit{
$^\S$Aachen Institute for Advanced Study in Computational Engineering Science (AICES), \\
RWTH Aachen University, Templergraben 55, 52056 Aachen, Germany \\[1.1mm]
$^\dag$ Department of Mechanical Engineering, IIT Kanpur, Kanpur - 208016, India}}




\end{center}
\vspace{-4mm}

\renewcommand{\thefootnote}{\arabic{footnote}}

\begin{center}

\small{Published\footnote{This pdf is the personal version of an article whose journal version is available at \href{https://doi.org/10.1016/j.carbon.2019.12.014}{https:/\!/sciencedirect.com}} 
in \textit{Carbon}, \href{https://doi.org/10.1016/j.carbon.2019.12.014}{DOI: 10.1016/j.carbon.2019.12.014} \\
Submitted on 2 October 2019; Revised on 6 December 2019; Accepted on 7 December 2019} 

\end{center}

\vspace{-3mm}


\rule{\linewidth}{.15mm}
{\bf Abstract} \\
In this paper, the validity and accuracy of three interatomic potentials and the continuum shell model of \citet{Ghaffari2018a} are investigated. The mechanical behavior of single-layered graphene sheets (SLGSs) under uniaxial stretching, biaxial stretching and pure bending is studied for this comparison. The validity of the molecular and continuum models is assessed by direct comparison with density functional theory (DFT) data available in the literature. The molecular simulations are carried out employing the MM3, Tersoff and REBO+LJ potentials. The continuum formulation uses an anisotropic hyperelastic material model in the framework of the geometrically exact Kirchhoff-Love shell theory and isogeometric finite elements.  Results from the continuum model are in good agreement with those from DFT. The results from the MM3 potential agree well up to the point of material instability, whereas those from the REBO+LJ and Tersoff potentials agree only for small deformations. Only the Tersoff potential is found to yield auxetic response in SLGSs under uniaxial stretch. Additionally, the transverse vibration frequencies of a pre-stretched graphene sheet and a carbon nanocone are obtained using the continuum model and molecular simulations with the MM3 potential. The variations of the frequencies from these approaches agree within an error of $\approx5\%$.

\vspace{-4mm}
\hspace{-3.5mm}
\rule{\linewidth}{.15mm}
\section{Introduction} \label{Introduction}
Graphene is an atom-thick two-dimensional (2D) hexagonal lattice of covalently bonded carbon atoms, which can be exfoliated from bulk graphite \citep{Novoselov2004}. Due to its excellent mechanical \citep{Lee2008a}, electrical and thermal properties \citep{Tang2009}, it has many industrial applications in the fields of nanocomposites \citep{Stankovich2006}, nano-electromechanical systems \citep{Bunch2007} and electronic devices \citep{Westervelt2008}. The electronic properties, in particular the band gap of a single-layered graphene sheet (SLGS) can be altered by applying uniaxial or biaxial strains \citep{ Guinea2009, liu2017, Ding2010} on it. This method is popularly called as \textit{strain engineering}. Also, uniaxial or biaxial strain in an SLGS of a bilayered graphene sheets (BLGSs) leads to change in the stacking dislocations that commence changing electronic or optical properties \citep{Lebedeva2019a}. However, successful operation of such strain-tunable devices will hinge on the satisfactory mechanical response of SLGS under applied strain field, small or large.  
\\
Another carbon based nanostructure obtained due to pentagonal and heptagonal defects in graphene is carbon nanocone (CNC) \citep{Ge1994, Liu2010}. CNCs have potential applications in the scanning probe microscopy \citep{Charlier2001}, field emission electron source \citep{Adisa2011} and molecular pumps \citep{xuz2016}. 

The mechanical response, static or dynamic, of a SLGS/CNC under imposed strains can be studied comprehensively using quantum-mechanical or molecular statics/dynamics (MS/MD) simulations. Alternatively, one can develop a suitable equivalent continuum model of the SLGS/CNC which reproduces identical response as obtained from these simulations for a large range of applied strain field. We note that the quantum-mechanical simulations of CNCs will be computationally inefficient, however, a continuum membrane model calibrated using quantum-mechanical or MS/MD simulations based data can be employed to study deformations/dynamics of CNCs. 

In the following, we begin with surveying the literature reporting linear response of SLGSs and the material constants derived from it. In Table~\ref{t:elastic}, the elastic moduli of a SLGS along the armchair, $E_\text{AC}$, and zigzag, $E_\text{ZZ}$, directions obtained from these methods are reported. As seen, some authors report different stiffness in the armchair and zigzag directions. This table also reveals ambiguity in the literature regarding the isotropic behavior of SLGS, even in the small deformation regime. Subsequently, in this paper, relevant discussion on this aspect from our findings are reported.\\
\begin{table}[h]
\footnotesize
\centering
\caption{Elastic properties of graphene reported in the literature. NA = Not available. \label{t:elastic}}
\begin{tabular}{lcccccc}
  \hline
   Ref.  & Year & Method/potential & $E_{\mathrm{AC}}$ [TPa]  & $E_{\mathrm{ZZ}}$ [TPa] & Thickness [nm]  \\ [1mm]
  \hline
\citep{VanLier2000}  & 2000 & \textit{Ab-initio} & 1.100 & 1.100 & 0.340 \\
\citep{Reddy2006} & 2006 & Molecular structural mechanics & 1.096--1.125 & 1.106--1.201 & 0.340 \\
     & & (Tersoff-Brenner) &  \\
\citep{Konstantinova2006} & 2006 & DFT & 1.250 & 1.250 & 0.340 \\

\citep{Liu2007}  &  2007 & \textit{Ab-initio} & 1.050 & 1.050 & 0.334 \\
\citep{Lee2008a} & 2008 & Experimental (nano indentation) & 1 $\pm $ 0.100 & 1 $\pm $ 0.100  & 0.335 \\
\citep{Faccio2009} & 2009 & DFT & 0.964  & 0.964 & 0.340 \\
\citep{Sakhaee-Pour2009} & 2009 &  Molecular structural mechanics & 1.040 & 1.042 & 0.340 \\
\citep{Gao2009} & 2009 & Quantum molecular dynamics & 1.100 & 0.600 & 0.335 \\
\citep{Zhao2009} & 2009 & Orthogonal tight-binding and MD & 1.01 $\pm $ 0.030 & 1.01 $\pm $ 0.030  & 0.335 \\
\citep{Scarpa2009} & 2009 & Truss-type analytical models & \\
 &  & with AMBER & 1.378 & 1.303 & NA  \\
 &  & with MORSE & 1.379 & 1.957 & NA \\
\citep{Pei2010} & 2010 & MD (AIREBO) & 0.890 & 0.830 & NA \\
\citep{Georgantzinos2010} & 2010 & Molecular structural mechanics & 0.721 & 0.737 & 0.340 \\
\citep{Ni2010} & 2010 & MD (Tersoff) & 1.130 & 1.050 & 0.335 \\
\citep{Gupta2010a} & 2010  & Molecular statics (MM3) & 3.380 & 3.400 & 0.100 \\
\citep{Liu2012}  & 2012 & MD (AIREBO) & 1.097 & 0.961 & 0.335\\
\citep{Kalosakas2013} & 2013 & Molecular structural mechanics and MD   & 1.070   & 1.070  & 0.335 \\
\citep{Alzebdeh2014} & 2014 & Molecular structural mechanics & 1.100 &1.100  & 0.34& \\
& & (Modified MORSE) & & & & \\
\citep{Genoese2017} & 2017 & Space frame approach & & & & \\
& & with AMBER & 0.780 & 0.819 & NA \\
& & with MORSE & 0.890 & 0.938 & NA \\
\citep{Singh2018a2} & 2018 & Multiscale model (MM3) & 0.927 & 0.927 & 0.335 \\
  \hline
\end{tabular}
\end{table}\noindent
In MD and MS simulations, the accuracy of the results depends on the potential defining atomic interactions. For carbon structures, popular potentials are: MM3 \citep{Allinger1989}, Tersoff \citep{Tersoff1989}, the first and second generation reactive empirical bond order (REBO) \citep{Brenner1990, Brenner2002}, adaptive intermolecular reactive empirical bond order (AIREBO) \citep{Stuart2000}, and ReaxFF \citep{Chenoweth2008}. Recently, \citet{Lebedeva2019} investigated the applicability of these potentials (except MM3) up to 3$\%$ uniaxial elongation and reported that the considered potentials fail to reproduce precisely the experimental and \textit{ab-initio} in-plane and out-of-plane deformations of a SLGS. Employing the MM3 potential in MS simulations, \citet{Gupta2010a} reported significant increase of the transverse vibration frequencies of SLGS under pre-stretch. Similarly, \citet{liao2011} studied the influence of temperature, cone height, and cone angles on the mechanical behavior of CNCs under uniaxial tensile and compression employing the Tersoff potential.\\
Some equivalent continuum structures have been proposed in the linear regime to study vibrations of SLGSs \citep{Kitipornchai2005, Jiang2014, Sakhaee-Pour2009, Georgantzinos2010, Chowdhury2011, Atalaya2008, Mai2012} and CNCs \citep{Seyyed2012, Hu2012}. Using a continuum plate model, \citet{Jiang2014} investigated the effect of size, shape and boundary conditions on vibration of SLGSs and multi-layered graphene sheets (MLGSs). Apart from the elastic properties, the vibrational behavior of SLGSs has also been studied using molecular structural mechanics \citep{Sakhaee-Pour2008,sadeghi2010}. \citet{sadeghi2010} studied the nonlinear vibrations of SLGS and reported that the fundamental frequency depends on the amplitude of vibration and length of SLGS. Using MS based on the universal force field (UFF) model, \citet{Chowdhury2011} reported that the natural frequencies of SLGSs are insensitive to the chirality. \citet{Singh2018} used a multiscale method to study the effect of pre-tension on the nonlinear static and dynamic response of SLGSs. They have reported that the pre-tension significantly increases natural frequencies. \citet{Singh2015} also studied the nonlinear elastic response of SLGSs and reported that SLGSs show softening behavior at small strains and hardening behavior at large bending. \citet{LuQ2009} studied the uniaxial in-plane and bending deformations of SLGS using molecular simulations employing REBO potential and equivalent continuum model. They have reported that SLGS exhibits nonlinear and anisotropic response under finite axial stretch. \\
There are some studies on vibration of nanocones which are in their relaxed state. For example, \citet{Seyyed2012} studied the vibrational properties of CNCs of different heights and cone angles using MS simulations and reported that the transverse vibrational frequencies reduce with increase in cone angle and height. \citet{Hu2012} modeled CNCs as tapered beams using the Euler-Bernoulli and Timoshenko beam theory to study transverse vibrations.\\
In this paper, the nonlinear mechanical response of a square graphene sheet under uniaxial and biaxial in-plane stretch is studied using MS/MD, DFT and a hyperelastic continuum model based on DFT data \citep{Shirazian2018_01}. The second generation REBO+LJ and Tersoff potentials are used in the MD simulations, and the MM3 potential is used in the MS simulations. For the Tersoff potential, two sets of parameterization are considered: the original parameters proposed by \citet{Tersoff1989} and the modified parameters proposed by \citet{Albe2005}. These two parameterizations are denoted ``Tersoff" and ``modified Tersoff", respectively, henceforth. The validity of the interatomic potentials is investigated in the linear and nonlinear deformation regime by comparing the results from MS/MD simulations with those obtained from DFT simulations \citep{Shirazian2018_01}. The frequencies of the transverse vibrations of the SLGS and a CNC under stretch, obtained from MS simulations, are compared with those obtained from the continuum model.\\
The remainder of this paper is arranged as follows: Section~\ref{Molecular simulations} describes the molecular simulation methods and the interatomic potentials used in the present study. A brief introduction to the continuum model is given in Section~\ref{Equivalent continuum model}. Numerical results are then presented in Section~\ref{Numerical results} followed by conclusions in Section~\ref{Conclusions}.
\section{Molecular simulations}\label{Molecular simulations}
This section presents details about molecular simulations and the mathematical expressions of the interatomic potentials considered.\\
Molecular simulations are carried out by solving the Newtonian equations of motion for the atoms,
\eqb{lll}\label{dyna}
\bF_{\!I}= m_{I}\,\ddot{\br}_{\!I}~,
\eqe
where $\br_{I}$ and $m_I$ are the position and mass of the atom~$I$, and $\bF_{I}$ is the interatomic force on atom~$I$. $\bF_{I}$  can be determined from the potential function $U\!\left(\text{\textbf{r}}^\textit{N}\right)$ as
\eqb{lll}\label{dyna}
\bF_{I}=-\ds \pa{U}{\br_{I}}~,
\eqe
where  $\mr^{N} \mathrel{\mathop:}=  \{\br_1, \br_2, \br_3, ..., \br_N\}$ are the positions of all the atoms.\\
The response of a graphene sheet under uniaxial and biaxial stretching, and bending can be computed using MS simulations for all the potentials discussed here. However, the computation of the bending stiffness based on plate vibration requires modal analysis. Whereas Tinker (coded with the MM3 potential) provides modal analysis, LAMMPS (coded with the REBO+LJ and Tersoff potentials) does not. Therefore, in Tinker the simulations are performed at 0 K, while in LAMMPS we perform MD at extremely low temperatures (0.1 K), that do not introduce entropic effects, and use fast Fourier transform to extract the vibration frequencies from the transverse time response of the atoms. \\
At 0 K, the average velocity fluctuations of the atoms are zero. These systems are thus static problems that can be solved by finding their minimum energy configurations. Here, this is done using the Broyden-Fletcher-Goldfarb-Shanno (BFGS) method \citep{nocedal}. Before applying any loads, the system under consideration (here SLGS/CNC) should itself be brought into the minimum energy configuration called the \textit{reference configuration}. We do this initial relaxation step by bringing the root mean square gradient of $U$ to 0.001 Kcal/mol/\AA. Subsequently, we apply the desired boundary conditions and compute the eigenvalues (frequencies) and eigenvectors (mode shapes) of the mass-weighted Hessian of the SLGS/CNCs at various stretch values using the \textit{Vibrate} subroutine of Tinker \citep{Ponder2004}.\\
In MD simulations, the specification of a finite temperature leads to the fluctuating velocity in the system. MD systems thus require transient solution approaches. Further, they need to be thermally equilibrated in order to reach quasi-static states at macro-scales. However, first the relaxed or minimum energy configuration of the system should be obtained. This is done here with the Polak-Ribiere’s conjugate gradient method \citep{Polak1969} in a quasi-static approach with absolute zero temperature and zero velocity. The energy\footnote{LAMMPS uses normalized energy units for the energy minimization criterion, defined as the relative change of the energy between two successive iterations divided by the energy magnitude at the current iteration.} and force tolerances for terminating the energy minimization are set to $10^{-8}$ and $10^{-8}$ eV/\AA, \ respectively. Subsequently, the system is thermally equilibrated at constant volume and temperature of 0.1 K for 50 ps with a timestep of 1 fs to obtain the reference configuration. The temperature is maintained at 0.1 K employing a Nose-Hoover thermostat  \citep{Evans1985} with three chains during deformations. The MD simulations are performed with the Large-scale Atomic/Molecular Massively Parallel Simulator (LAMMPS) \citep{lammps}.\\
In molecular simulations at approximately 0 K, the virial stress is defined by \citep{Tsai1979,Swenson1983}
\eqb{lll} \label{virial}
\bar{\sigma}_{I}^{ij}\dis \ds \frac{1}{V_{I}} \ds  \sum_{J =1}^{N_p} f_{IJ}^{i} \,r_{IJ}^{j}~,
\eqe
where $V_I$ is the volume occupied by atom~$I$, and $i$ and $j$ denote the Cartesian components along the \textit{x}, \textit{y} and \textit{z} directions, respectively. $N_p$ is the number of neighboring atoms of atom~$I$, $f_{IJ}^i$ is the force on atom~$I$ due to atom~$J$, and $ r_{IJ}$ is the distance between atom~$I$ and $J$. For 2D materials, such as graphene, it is natural to introduce the stress as force per length. Given (\ref{virial}), this stress follows  as
\eqb{lll}\label{eq:12}
\sigma_I^{ij}\dis \ds \ds\frac{V_I}{A_I} \bar\sig_I^{ij}~,
\eqe
where $A_I$ is the sheet area attributed to atom $I$. Through definition Eq.~(\ref{eq:12}), the usage of a thickness for SLGSs is avoided. \\
The interatomic potentials are described in the following subsections.
\subsection{MM3 Potential}
The MM3 potential consists of the first and higher order expansions of bond stretching, angle bending, and torsion. This potential also incorporates the cross terms\footnote{The cross term between the bending and torsion is not considered.} among the mentioned contributions and between angle bending and out-of-plane bending \citep{Allinger1989}. The expression of the MM3 potential is given by \citep{Allinger1989}
\eqb{lll} \label{eq:mm3}
U_{\text{MM3}}= \ds \sum_{I} \sum_J \big(U_{\text{s}}+U_{\theta}+U_{\phi}+U_{\text{s}\theta}+U_{\theta \theta^{'}}+U_{\phi \text{s}}\big)+\sum_I \sum_K U_{\text{vdW}}~,
\eqe
where $U_\text{s}$, $U_{\theta}$ and $U_\phi$ are the energy contributions corresponding to changes in bond length, bond angle and dihedral angle, respectively. $U_{\text{s} \theta}$, $U_{\theta \theta^{'}}$ and  $U_{\phi \text{s}}$ account for energies of the cross-term interactions between bond stretch and angle bending, angle bending and out-of-plane-bending, and bond stretch and dihedral angle, respectively. $U_{\text{vdW}}$ defines van der Waals (vdW) attraction and steric repulsion in the form $({r_\text{c}}/r_{IJ})^6$ and $\exp(-12r_{IJ}/r_\text{c}) $, where $r_c$ is the cut-off distance. Further details of these terms are given in Appendix \ref{mm3-1}.
\subsection{REBO+LJ Potential}
The REBO+LJ potential consists of two parts. The covalent bonds between carbon atoms are modeled using the second generation REBO potential, which is widely used for the formation and breaking of bonds in carbon structures. The REBO part of the potential is \citep{Stuart2000} 
\eqb{lll} \label{eq:rebo}
U_{\text{REBO}}=\ds \sum_I \sum_{J=I+1}\big[E_{\text{R}}(r_{IJ})+b_{IJ}\,E_{\text{A}}(r_{IJ})\big]~,
\eqe
where $r_{IJ} $ is the distance between a pair of atoms and $b_{IJ}$ is an empirical bond-order term. $E_{\text{R}}$ and $E_{\text{A}}$ are, respectively, the repulsive and attractive terms and are given in Appendix \ref{reb-1}.
The vdW attraction and steric repulsion are modeled by the standard 12-6 Lennard-Jones (LJ) potential \citep{Lj1924}
\eqb{lll}
U_{\text{LJ}} =\ds 4 \epsilon \left[ \left(\frac{\sigma}{r_{IJ}}\right)^{12} - \left(\frac{\sigma}{r_{IJ}}\right)^6 \right]~,
\eqe
where $\sigma $ and $\epsilon $ are the LJ parameters. The vdW energy is only included when the covalent bond energy from the REBO potential become zero, i.e., after breakage of the covalent bond.
\subsection{Tersoff Potential}
The Tersoff potential is a pair-like potential in which the strength of a bond depends on the local environment, i.e., an atom with fewer neighboring atoms forms a stronger bond than the atom with more neighboring atoms. In the literature, the mechancial behavior of carbon structures has been studied successfully using the Tersoff \citep{Tan2013, Buda2015} and modified Tersoff \citep{Win2018} potentials. Both are described by \citep{Tersoff1989, Albe2005}
\eqb{lll} \label{eq:tersoff}
U_{\text{Tersoff}}=\ds \frac{1}{2}\ds \sum_I \sum_{J\neq I}f_\text{c}(r_{IJ})\big[f_{\text{R}}(r_{IJ})+b_{IJ}\,f_{\text{A}}(r_{IJ})\big]~,
\eqe
where $r_{IJ} $ is the distance between the atom pairs $I$ and $J$, $b_{IJ}$ is the bond order term, $ f_{\text{R}} $ and $ f_{\text{A}}$ are the repulsive and attractive terms, respectively, and $ f_{\text{c}}$ is a smooth cutoff function.
Details of the potential are given in Appendix \ref{ter-1}.
\section{Continuum model} \label{Equivalent continuum model}
As a homogenized structure, the SLGS and CNC are modeled based on the shell formulation of \citet{Duong2017a} and the anisotropic hyperelastic material model of \citet{Ghaffari2018a}. \citet{Ghaffari2018a} formulated the strain energy density, per unit area of the initial configuration, based on a set of invariants $\sJ_{i}$ , i.e. \citep{Ghaffari2018c,Ghaffari2018a}
\eqb{l}
W(\sJ_{1},\sJ_{2},\sJ_{3}) = W_{\text{m}}^{\mathrm{dil}}(\sJ_{1}) +W_{\text{m}}^{\mathrm{dev}}(\sJ_{2},\,\sJ_{3};\sJ_{1})+W_{\text{b}}(\kappa_{1},\kappa_{2};\sJ_{1})~,
\label{e:total_strain_energy}
\eqe
where $W_{\text{m}}^{\mathrm{dil}}$ and $W_{\text{m}}^{\mathrm{dev}}$ are the pure dilatational and deviatoric parts of the membrane strain energy density, respectively, and $W_{\text{b}}$ is the bending strain energy density.
These terms are defined as
\eqb{lll}
 W^{\mathrm{dil}}_{\mathrm{m}} \dis \varepsilon\big[1 - (1+\hat{\alpha}\,\sJ_{1})\,\exp(-\hat{\alpha}\,\sJ_{1})\big]~,
 \eqe
 \eqb{lll}
 W^{\mathrm{dev}}_{\mathrm{m}} \dis 2\,\mu(\sJ_{1})\,\sJ_2 + \eta(\sJ_{1})\,\sJ_3~,
\eqe
\eqb{lll} \label{eq:b_c}
\ds W_{\mathrm{b}} \dis \ds J\frac{c_{\text{b}}}{2}\left(\kappa_1^2+\kappa_2^2\right)~,
\eqe
where $\mu$ and $\eta$ are defined as
\eqb{lll}
\mu(\sJ_{1}) \dis \mu_0 - \mu_1\,e^{\hat{\beta}\,\sJ_{1}}~,\\[2mm]
\eta(\sJ_{1}) \dis \eta_0 - \eta_1\,\sJ_{1}^2~.
\eqe
The material constants $\varepsilon$, $\hat{\alpha}$, $\mu_0$, $\mu_1$, $\hat{\beta}$, $\eta_0$, $\eta_1$, and $c_{\text{b}}$ from \citet{Shirazian2018_01} are used in the continuum model, see Tables \ref{t:Graphene_cons} and \ref{t:graphene_bending_material_cons}. 
\begin{table}[h]
  \centering
     \caption{Membrane behavior: Hyperelastic material constants determined by fitting to DFT calculations based on generalized gradient approximation (GGA) according to \citet{Kumar2014_01} and \citet{Shirazian2018_01}.}\label{t:Graphene_cons}
    \begin{tabular}{c c c c c c c c }
      \hline
        & $\hat{\alpha}$ & $\varepsilon~[\textnormal{N/m}]$  &  $\mu_0~[\textnormal{N/m}]$ & $\mu_1~ [\textnormal{N/m}]$ & $\hat{\beta}$ & $\eta_0~[\textnormal{N/m}]$ & $\eta_1~[\textnormal{N/m}]$ \\
        \hline
      \citet{Kumar2014_01}  & 1.53 & 93.84  & 172.18 & 27.03 & 5.16 & 94.65 & 4393.26 \\
      \citet{Shirazian2018_01}  & 1.435 & 103.9 & 182.6 & 34.94 & 4.665 & 83.46& 3932  \\
      \hline
    \end{tabular}
\end{table}
\begin{table}
  \centering
  \caption{Bending stiffness,     $c_{\text{b}}$ in [nN$\cdot$nm] (in [eV]), according to various atomistic models. FGBP = first generation Brenner potential; SGBP = second generation Brenner potential; QM = quantum mechanics.}
  \label{t:graphene_bending_material_cons}
     \begin{tabular}{c c c c c}
     \hline
     FGBP \citet{Lu2009_01} & SGBP \citet{Lu2009_01} & QM \citet{Kudin2001_01} & \citet{Shirazian2018_01}\\
     \hline
0.133 (0.83)& 0.225 (1.40) & 0.238 (1.49) & 0.238 (1.49)\\
    \hline
  \end{tabular}
\end{table}$J$ is the surface area change, and $\kappa_{1}$ and $\kappa_{2}$ are the principal surface curvatures \citep{Ghaffari2019}. $\sJ_{1}$ and $\sJ_{2}$ capture isotropic dilatation and shear deformation, respectively, while $\sJ_{3}$ captures anisotropic shear deformation. $\sJ_{i}$ are given by
\eqb{lll}
\sJ_{1} \, \dis \, \ln J = \ln (\lambda_{1}\lambda_2)~,\\[2mm]
\ds \sJ_{2} \dis \ds \, \frac{1}{4}\left(\frac{\lambda^2_1}{\lambda^2_2}+\frac{\lambda^2_2}{\lambda^2_1}-2\right)~,\\[2mm]
\ds \sJ_{3} \dis \ds \, \frac{1}{8}\left(\frac{\lambda_1}{\lambda_2}-\frac{\lambda_2}{\lambda_1}\right)^3\,\cos(6\theta)~.
\eqe
$\lambda_1$ and $\lambda_2$ are the principal surface stretches with $\lambda_1\geq\lambda_2$. $\theta$ is the maximum stretch angle relative to the armchair direction and defined as
\eqb{lll}
\theta \, \dis \, \arccos{(\bY_{\!1}\cdot\hat{\bx})}~,
\eqe
where $\bY_{\!1}$ is the direction of the maximum stretch, see \citet{Ghaffari2018c} and \citet{Ghaffari2018a} for details. The material has isotropic behavior under pure dilatation, and anisotropic behavior only appears under large shear deformation. Material model (\ref{e:total_strain_energy}) is implemented in the nonlinear finite shell element formulation of \citet{Duong2017a}. The discretized weak form can be written as \citep{Duong2017a}
\eqb{lll}
\mM\,\ddot{\muu} +\mf_{\text{int}}(\muu)=\mf_{\text{ext}}(\muu)~,
\label{e:weak_form_dyn0}
\eqe
where $\mM$ is the mass matrix (see \citet{Ghaffari2018b} for the mass matrix of graphene), $\muu$ is the displacement vector and $\mf_{\text{int}}$ and $\mf_{\text{ext}}$ are the internal and external force vectors, respectively. They are assembled from the elemental contributions $\mf_{\text{int}}^{e}$ and $\mf_{\text{ext}}^{e}$, respectively, using the standard finite element assembly procedure. $\mf_{\text{int}}^{e}$ and $\mf_{\text{ext}}^{e}$ are defined by
\eqb{lll}
\mf_{\text{int}}^{e} \dis \ds \int\limits_{\Omega_{0}^{e} }{\tau^{\alpha\beta}\,\mN_{,\alpha}^{\text{T}}\,\ba_{\beta}~\dif A}+ \int\limits_{\Omega_{0}^{e} }{M_0^{\alpha\beta}\,{\mN}_{;\alpha\beta}^{\text{T}}\,\bn~\dif A}~,
\eqe
\eqb{lll}
\mf_{\text{ext}}^{e} \is \ds \int\limits_{\partial_t \Omega^{e} }{\mN^{\text{T}}\,\bt~\dif s}~,
\eqe
for the special case of zero body forces and zero boundary moments. Here $\Omega_{0}^{e}$ and $\Omega^{e}$ denote element domains in the reference and current configuration, respectively, and along $\partial_{\bt} \Omega$, the boundary traction $\bt$ is applied. $\ba_{\alpha}$ are the covariant tangent vectors and $\bn$ is the normal vector to the shell surface. $\mN$, $\mN_{,\alpha}$ and $\mN_{;\alpha\beta}$ denote the shape function matrix and its first parametric derivative and second covariant derivative \citep{Duong2017a}, and $(\bullet)^{\text{T}}$ is the transpose operator. The contra-variant components of the surface Kirchhoff stress tensor and the moment tensor are given by \citep{Duong2017a}
\eqb{lll}
\ds  \tauab \, \is \ds \, \pa{W}{\auab}~,
\eqe
\eqb{lll}
\ds M^{\alpha\beta}_{0} \, \is \ds \, \pa{W}{\buab}~,
\eqe
where $\auab$ and $\buab$ are the surface metric and curvature tensor components. The Cauchy and Kirchhoff stress components are connected by
\eqb{lll}
\sigab=\ds \frac{1}{J}\tauab~.
\eqe
Using a Taylor expansion about $\widehat{\muu}$ such that $\muu=\widehat{\muu}+\Delta\muu$, the linearized equations of motion become
\eqb{lll}
\mM\,\Delta\ddot{\muu} +\mK\,\Delta\muu =-(\mf_{\text{int}}+\mM\,\ddot{\widehat{\muu}}-\mf_{\text{ext}})~,
\label{e:weak_form_dyn}
\eqe
where $\mK:=\partial(\mf_{\text{int}}-\mf_{\text{ext}})/\partial\muu$ is the tangent stiffness matrix (see \citet{Ghaffari2018b} and \citet{Duong2017a} for details). For harmonic vibrations we have
\eqb{lll} \label{e:weak_form_dyn1}
\Delta \muu \, \dis \, \Delta \bar{\muu}\,e^{-i\omega t}~,
\eqe
where $\Delta \bar{\muu}$ are $\omega$ are the mode shape and frequency of the structure, respectively. Using \eqref{e:weak_form_dyn1}, Eq.~\eqref{e:weak_form_dyn} can be transformed to the standard eigenvalue problem
\eqb{lll}
\mK\,\Delta\bar{\muu}=\omega^2\,\mM\,\Delta\bar{\muu}~.
\label{e:eigen_prob}
\eqe
At each load increment, we solve Eq.~\eqref{e:weak_form_dyn0} iteratively using the Newton-Raphson method. Subsequently eigenvalue problem \eqref{e:eigen_prob} is solved at each load increment in order to obtain the variation of the frequencies with the loading.
\section{Numerical results} \label{Numerical results}
This section presents numerical results pertaining to the deformation and vibrations of a square SLGS and a CNC. The three molecular models of Sec. \ref{Molecular simulations} are compared with DFT data from \citet{Shirazian2018_01} and the DFT-based continuum model of Sec. \ref{Equivalent continuum model}. First, the minimum energy configuration of the SLGS is given in Subsection~\ref{min_en}. Then the procedure for the in-plane stretch of the SLGS is described in Subsection~\ref{procedure}, and the variation of strain energy and stresses is studied in Subsections~\ref{st_en} and \ref{stress}. The out-of-plane bending of SLGS is then discussed in Subsection~\ref{out_ben}, followed by the modal analysis of SLGS and CNC in Subsection~\ref{freq}. The transverse vibration frequencies of SLGS are calculated up to the instability point, where the ellipticity of the elasticity tensor is lost \citep{Kumar2014_01}. A pointwise summary of the main findings is finally given in Subsection~\ref{main_f}.
\subsection{Minimum energy configuration of SLGS} \label{min_en}
A SLGS with dimensions 10 nm $\times $ 10 nm and an initial bond length of 0.142 nm is considered. The mean and standard deviation of the bond length for the reference configuration employing the three potentials are shown in Table~\ref{mean_st}. In all cases, the equilibrium (i.e. mean) bond length is slightly different than the experimental value 0.1422 nm \citep{Bosak2007}. The difference can be attributed to the various functional forms of the potentials and the constants used therein. The deviations from the mean are due to the numerical errors that depend on the convergence tolerances considered for the energy minimization. Further, from Table~\ref{mean_st}, we note that after the minimization, the SLGS shrinks more when the MM3 potential is employed compared to REBO+LJ. On the other hand, the SLGS dilates after relaxation when the Tersoff potential is used.
\begin{table}[!htbp]
\footnotesize
\centering
\caption{The mean and standard deviation of the bond lengths for the reference configuration of the SLGS.} \label{mean_st}
\begin{tabular}[!htbp]{lcccc}
  \hline
    Potential & Mean [nm] &  Standard deviation [nm]   \\
    \hline
MM3 & 0.1345 & 0.00020  \\
REBO+LJ  & 0.1395 & 0.00036  \\
Tersoff & 0.1460 & 0.00028  \\
modified Tersoff & 0.1474 & 0.00026  \\
  \hline
\end{tabular}
\end{table}
\subsection{In-plane stretching of SLGS}
This section examines the elastic response of the square SLGS under uniaxial and biaxial stretch.
\subsubsection{Procedure} \label{procedure}
\begin{figure}
    \begin{subfigure}[p]{0.3\linewidth}
        \centering
     \includegraphics[width=50mm]{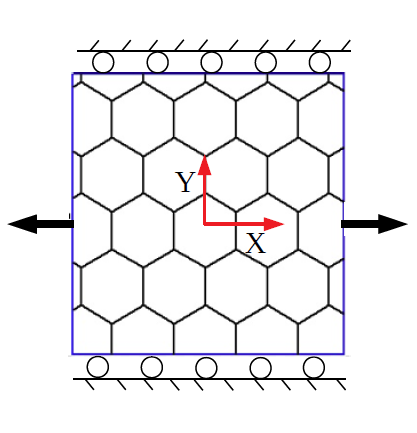}
         \vspace{-7mm}
        \subcaption{}
    \end{subfigure}
         \begin{subfigure}[p]{0.3\linewidth}
        \centering
 \includegraphics[width=50mm]{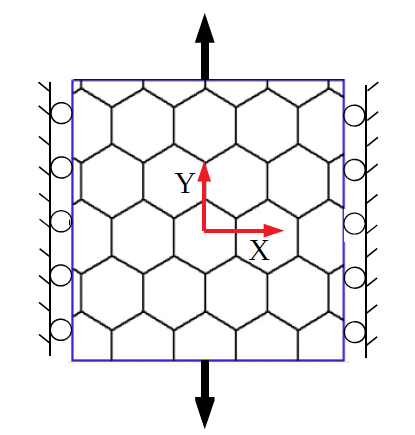}
         \vspace{-7mm}
        \subcaption{}
    \end{subfigure}
    \begin{subfigure}[p]{0.3\linewidth}
        \centering
    \includegraphics[width=50mm]{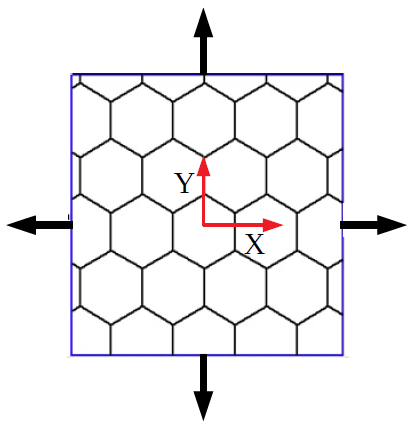}
        \vspace{-7mm}
        \subcaption{}
    \end{subfigure}
            \vspace{-3mm}
   \caption{Pre-stressed SLGS: Stretch along the (a) zigzag and (b) armchair direction, and (c) under pure dilatation.\label{fig:Gn_loading}}
\end{figure}

The MS simulations (with MM3) of the SLGS are performed with and without periodic boundary conditions (PBCs) to investigate the size effect.
In the absence of PBCs, the position in the current configuration ($x$, $y$) of the edge atoms is given by
\eqb{lll} \label{eq2}
\left[
  \begin{matrix}
  x \\
  y\\
  \end{matrix}
 \right] = \left[
  \begin{matrix}
  \lambda_{1} & 0 \\
  0 & \lambda_{2} \\
  \end{matrix}
   \right]
\left[
  \begin{matrix}
  X \\
  Y \\
  \end{matrix}
  \right],
\eqe
where $X$ and $Y$ are the initial position of the edge atoms in the reference configuration. For uniaxial stretch, $\lambda_{1}=\lambda$ and $\lambda_{2}=1$ or $\lambda_{1}=1$ and $\lambda_{2}=\lambda$ (i.e. the lateral direction is kept fixed), while for pure dilatation, $\lambda_{1}=\lambda_{2}=\lambda$ (see Fig.~\ref{fig:Gn_loading}). When PBCs are used, the periodic simulation box is deformed along $X$ and/or $Y$ with a stretch increment of 0.1 \AA . At each increment, the edge atoms are kept fixed and the system is relaxed to compute the potential energy and virial stress.
\begin{figure}[!htbp]
        \begin{subfigure}[t]{0.49\linewidth}
        \centering
 \includegraphics[height=70mm]{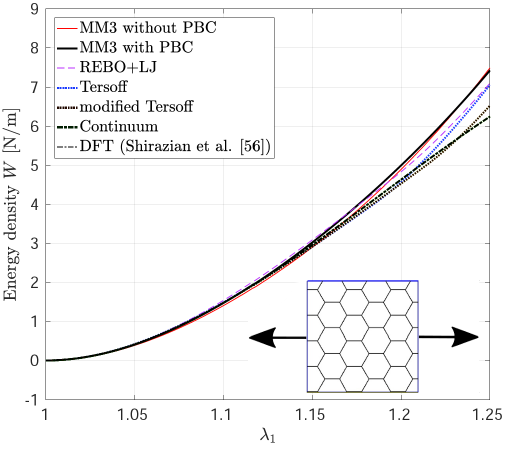}
         \vspace{-7mm}
        \subcaption*{(a)}
    \end{subfigure}
    \begin{subfigure}[t]{0.49\linewidth}
        \centering
 \includegraphics[height=70mm]{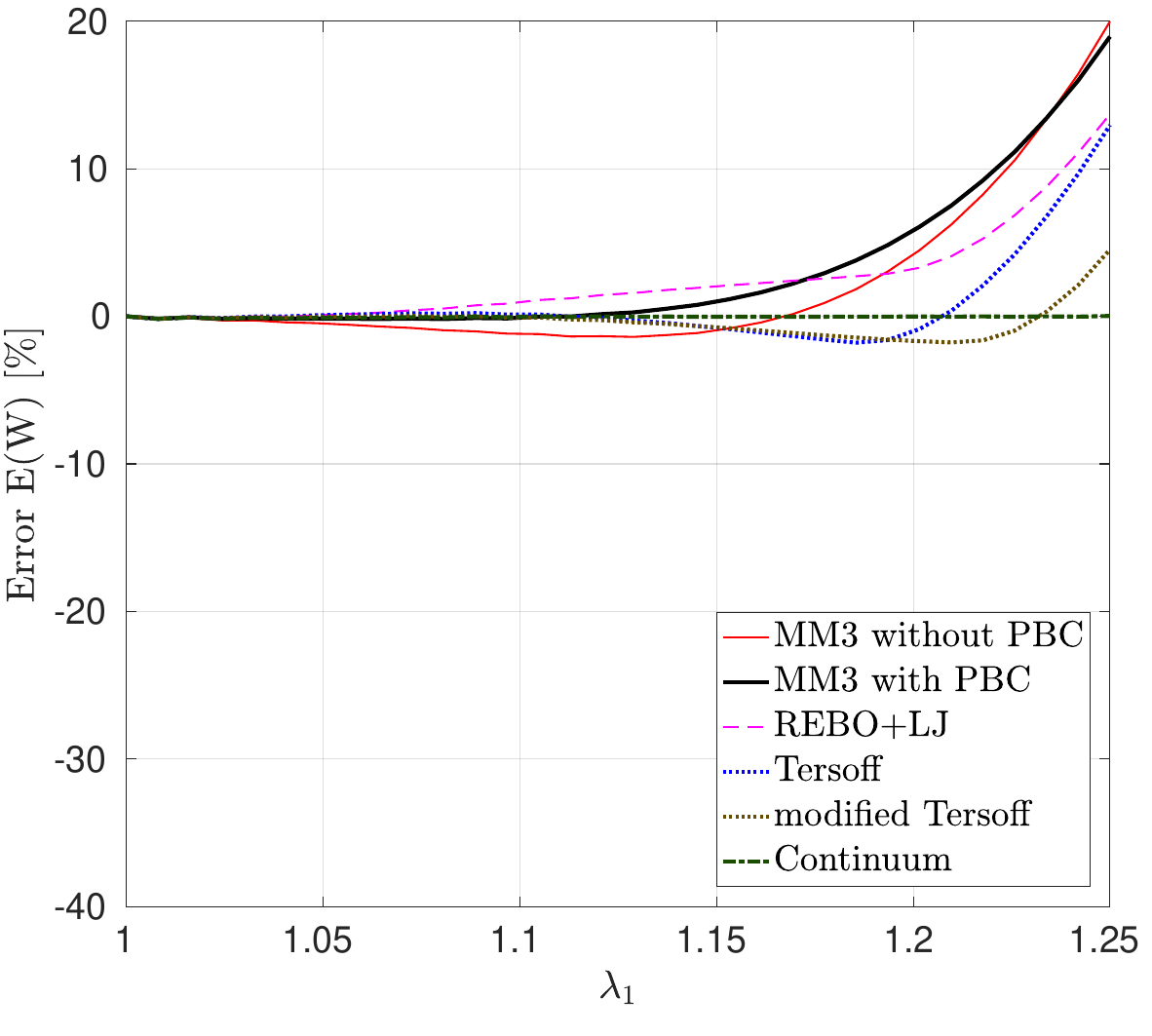}
         \vspace{-7mm}
        \subcaption*{(d)}
    \end{subfigure}
    \begin{subfigure}[t]{0.49\linewidth}
        \centering
    \includegraphics[height=70mm]{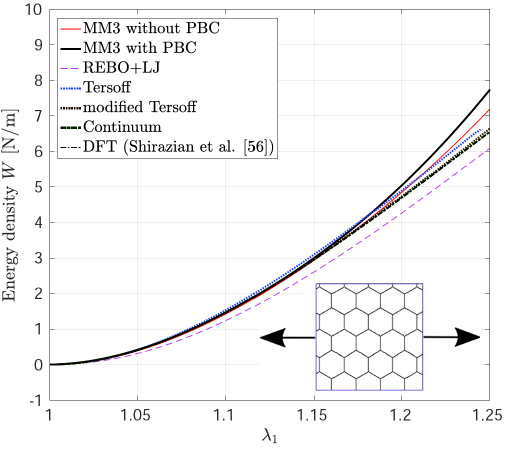}
        \vspace{-7mm}
        \subcaption*{(b)}
    \end{subfigure}
    \begin{subfigure}[t]{0.49\linewidth}
        \centering
    \includegraphics[height=70mm]{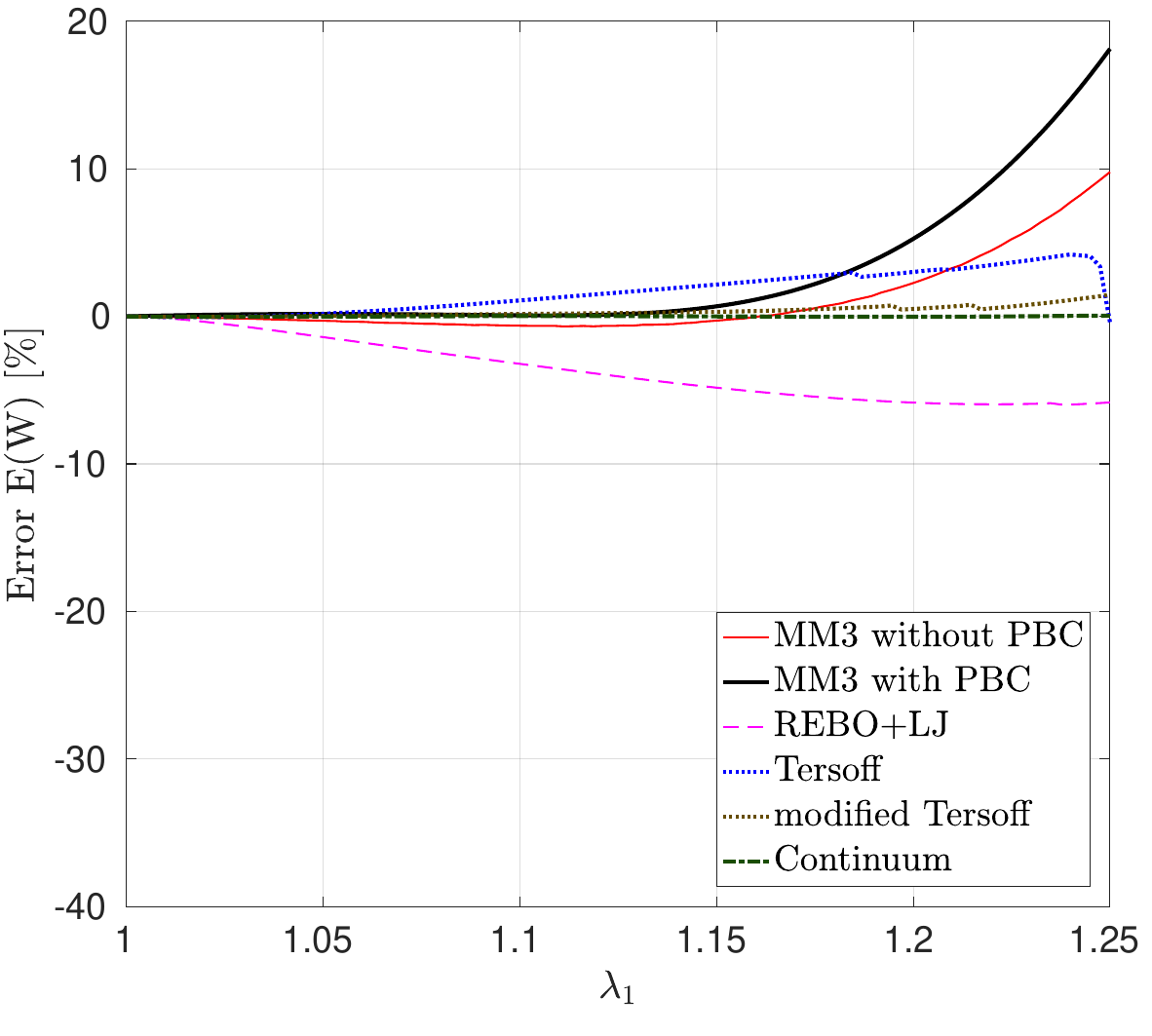}
        \vspace{-7mm}
        \subcaption*{(e)}
    \end{subfigure}
    \begin{subfigure}[t]{0.49\linewidth}
        \centering
     \includegraphics[height=70mm]{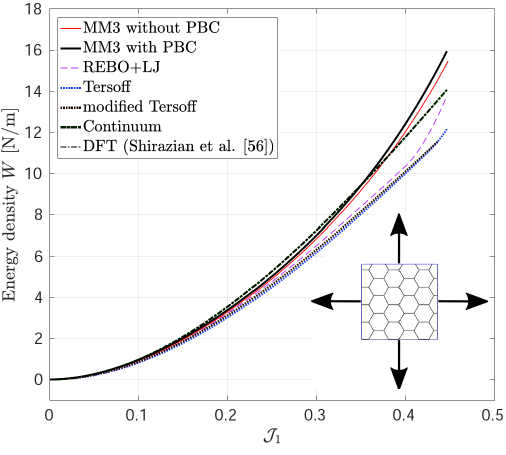}
         \vspace{-7mm}
        \subcaption*{(c)}
    \end{subfigure}
    \begin{subfigure}[t]{0.49\linewidth}
        \centering
     \includegraphics[height=70mm]{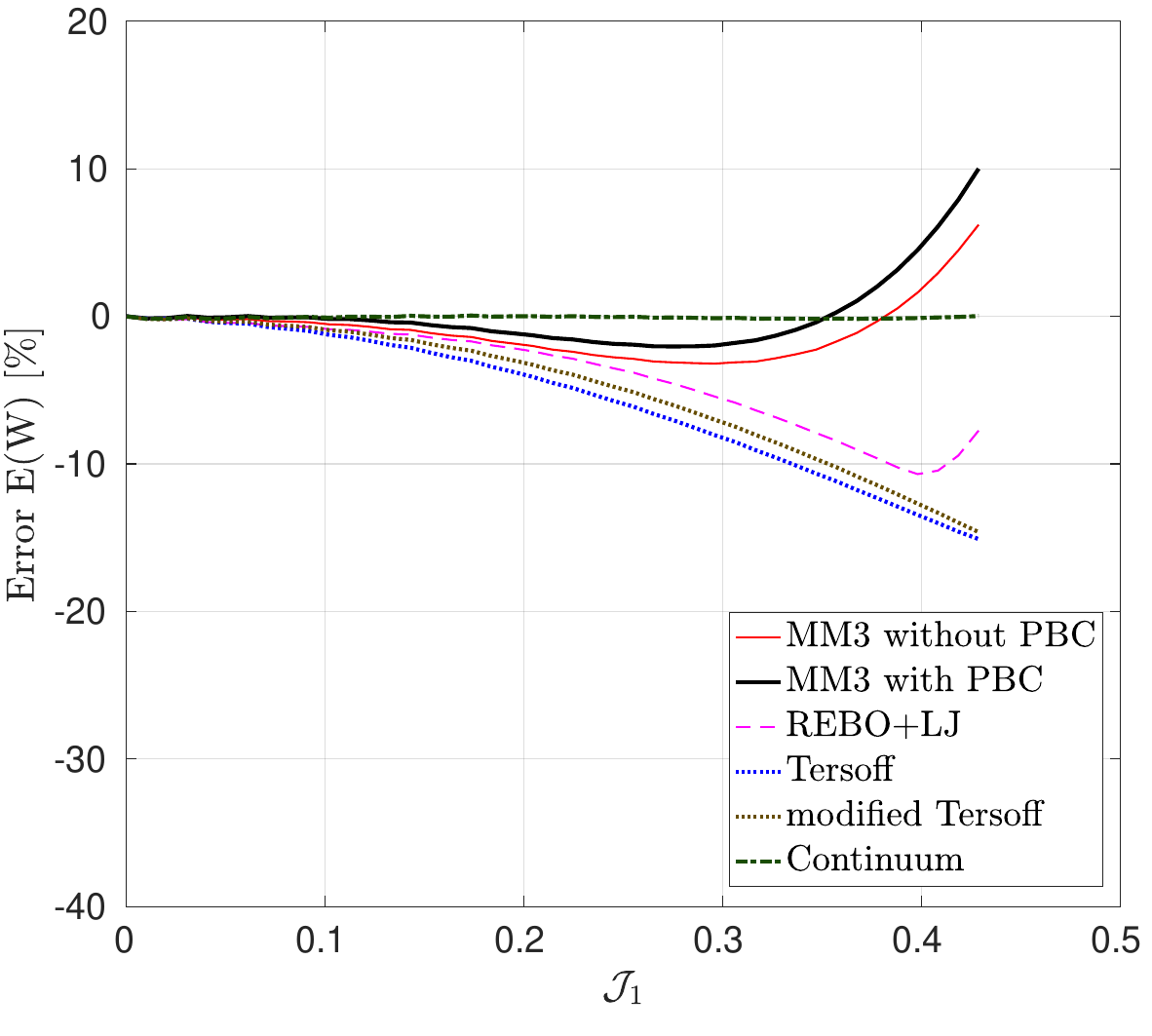}
         \vspace{-7mm}
        \subcaption*{(f)}
    \end{subfigure}
    \vspace{-1mm}
    \caption{Comparison of the total potential energy of a square SLGS under uniaxial stretch along (a) armchair, and (b) zigzag directions, and (c) in pure dilatation. The corresponding error $E(W)$ according to Eq.~(\ref{e:er_c}) is shown in (d), (e) and (f).\label{fig:energy_1}}
\end{figure} \\
In the MD simulations (with REBO+LJ and Tersoff), PBCs are employed. The SLGS is stretched uniaxially or biaxially using the deformation control method. A side of the periodic box ($l$) at any instant of time ($t$) is deformed according to $l(t) = l_0 (1 + e_{rate}t) $, where $l_0 = 10 \ \text{nm}$, is the initial length of the box and the applied strain rate is $e_{rate} = 0.001$/ps\footnote{This strain rate translates to the velocity 10 m/s for a SLGS with 10 nm length, which is much much less than the speed of sound wave propagation in SLGS, $\approx$ 21,000 m/s, and hence eliminates any transient effect.} \citep{lammps}. These parameters values are optimum to study the elastic behavior of the SLGS according to \citet{Zhao2009}.\\
In the continuum simulations, (\ref{eq2}) is applied quasi-statically without using PBCs.
\subsubsection{Strain energy}\label{st_en}
Figure~\ref{fig:energy_1} shows the potential energy variation versus the uniaxial and biaxial stretch computed from molecular simulations, the DFT-based continuum model and DFT \citep{Shirazian2018_01}. The relative error in the strain energy $W$ is shown in Fig.~\ref{fig:energy_1}d, \ref{fig:energy_1}e and \ref{fig:energy_1}f. The relative error in the quantity $X$ with respect to the DFT results is defined as
\eqb{lll} \label{e:er_c}
E(X) = \ds \frac{X - X_{\text{DFT}}}{\max(X_{\text{DFT}})}~,
\eqe
where $X$ results from the continuum and/or molecular simulations and $X_{\text{DFT}}$ is the DFT result. The maximum value is used to avoid division by zero. In MS simulations, it is found that the variation in the energy and stress (discussed subsequently) are almost identical with and without PBCs. These results confirm that the size effect on the elastic response is negligible when the diagonal length of the SLGS is over 10.0 nm as reported by \citet{Zhao2009}.\\
The strain energy results from the MM3 potential agree with the DFT results up to the stretch $\lambda_1 \approx 1.15$ for uniaxial loading and $ \mathcal{J}_1 \approx0.3$ for biaxial loading within $\approx5 \%$ error. The results from the REBO+LJ and Tersoff potentials agree within an error of $\approx $ 20$\%$. The continuum results are in excellent agreement with those from DFT simulations for the whole range under study (within $\approx 0.05\%$ error). This is due to the fact, that the continuum model has been calibrated directly from DFT data \citep{Shirazian2018_01}.
\subsubsection{Stresses} \label{stress}
\begin{figure}[h]
    \begin{subfigure}[t]{0.49\linewidth}
        \centering
 \includegraphics[height=65mm]{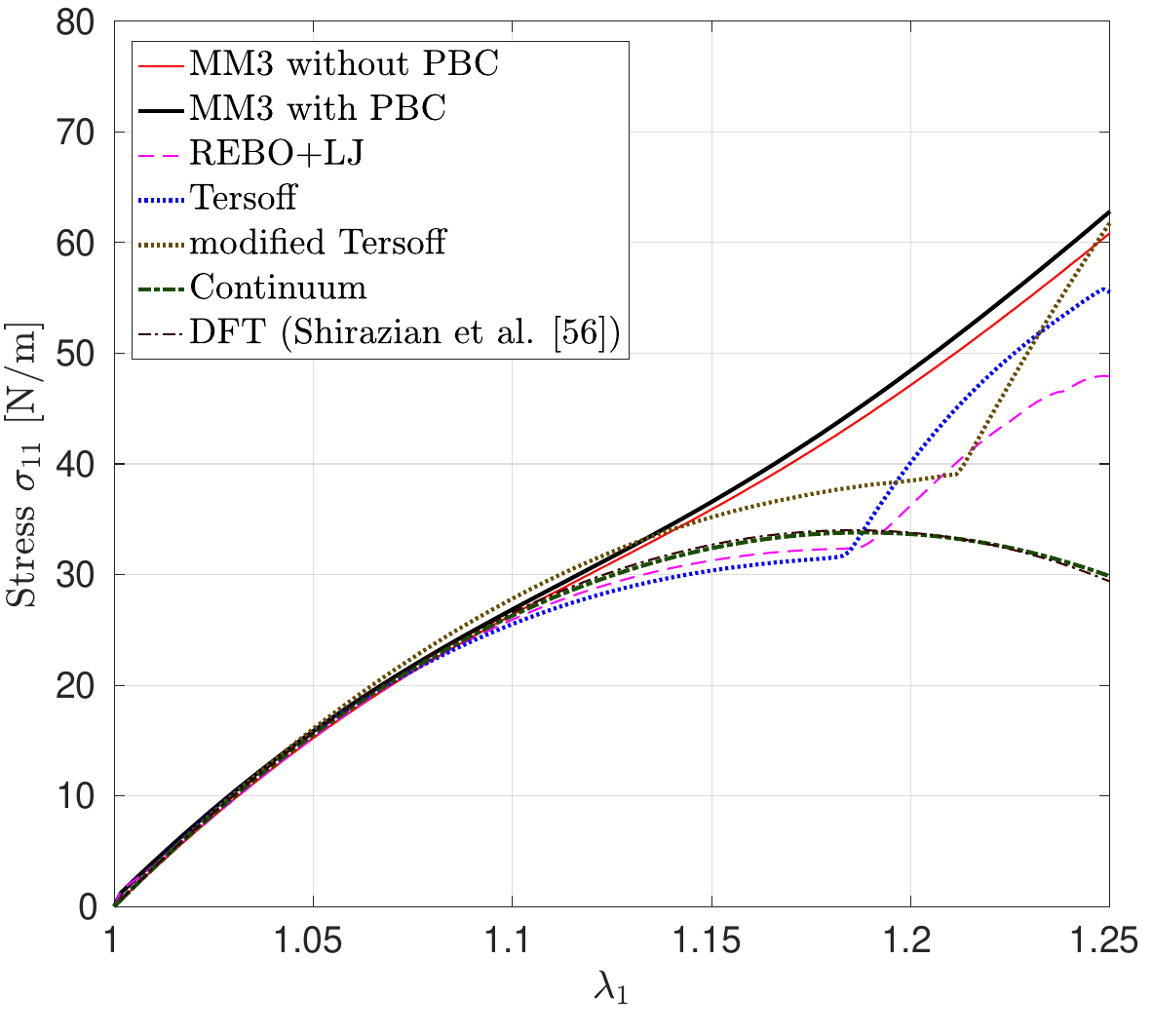}
         \vspace{-3mm}
        \subcaption{}
    \end{subfigure}
    \begin{subfigure}[t]{0.49\linewidth}
        \centering
    \includegraphics[height=65mm]{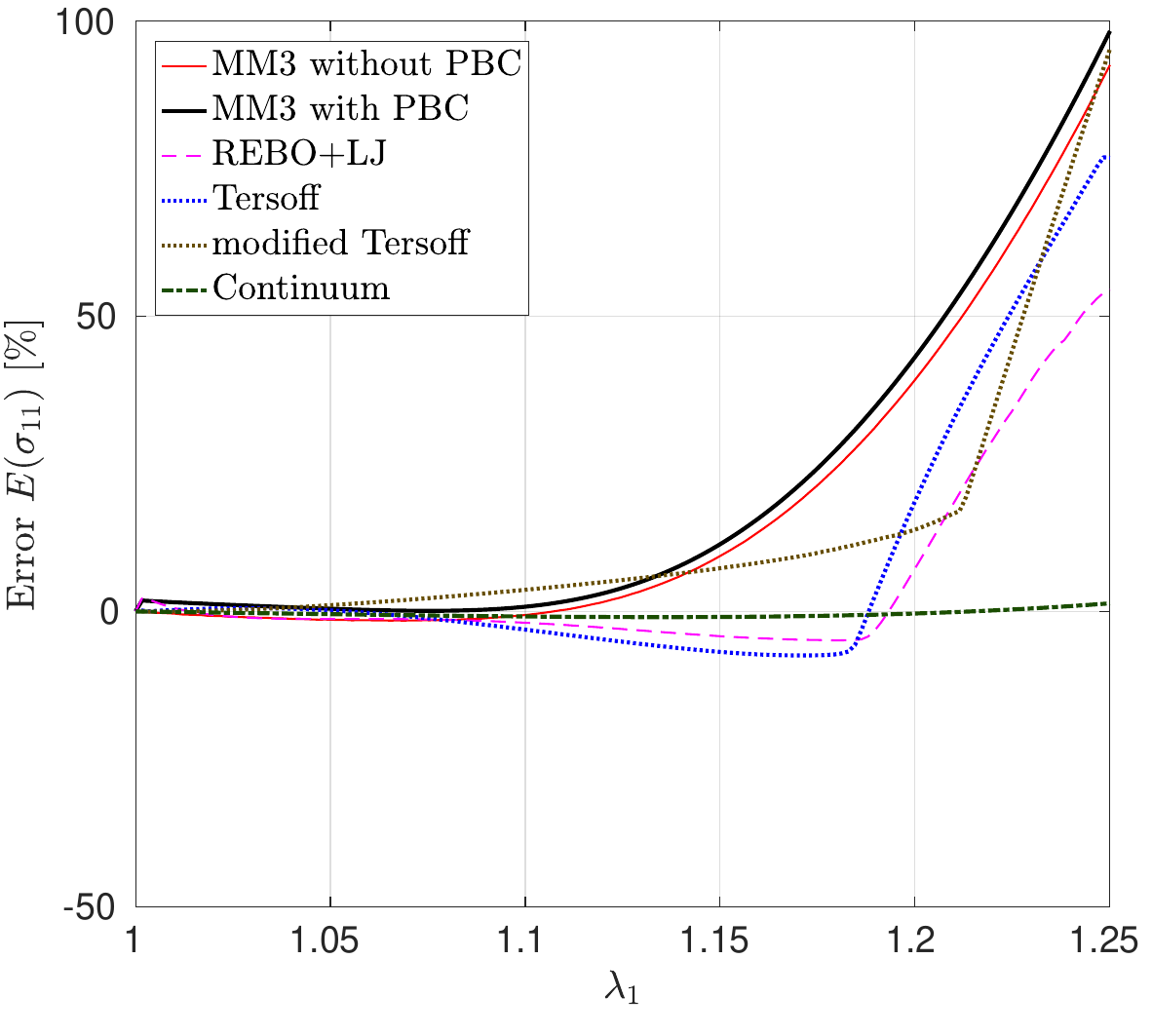}
        \vspace{-3mm}
        \subcaption{}
    \end{subfigure}
    \vspace{-3mm}
     \caption{Variation of (a) stress $\sigma_{11}$ and (b) error E($\sigma_{11}$) according to Eq.~(\ref{e:er_c}) as a function of stretch $\lambda_1$ in the armchair direction.\label{fig:long_stress_AC}}
\end{figure}

\begin{figure}[h]
    \begin{subfigure}[t]{0.49\linewidth}
        \centering
 \includegraphics[height=65mm]{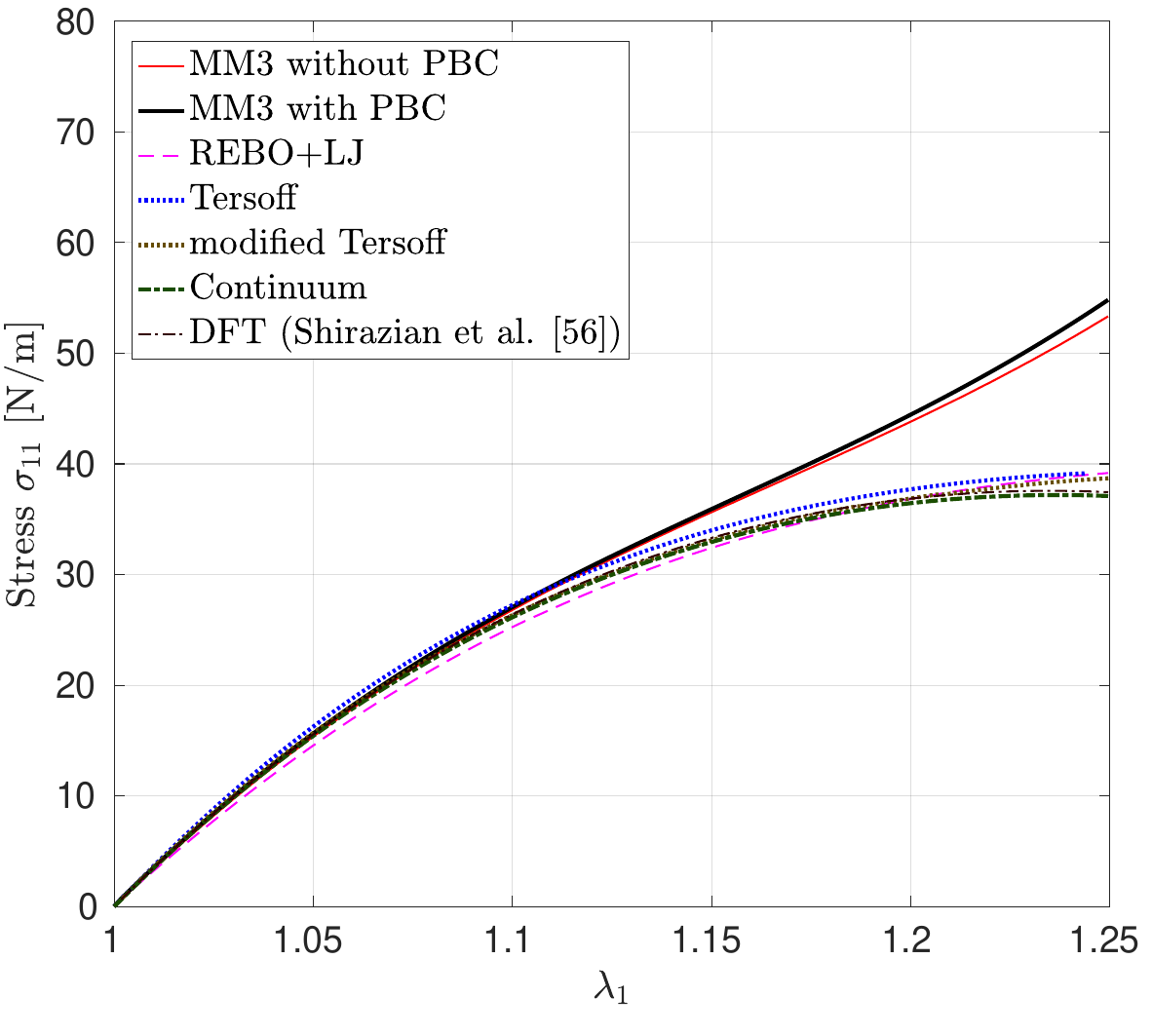}
         \vspace{-3mm}
        \subcaption{}
    \end{subfigure}
    \begin{subfigure}[t]{0.49\linewidth}
        \centering
    \includegraphics[height=65mm]{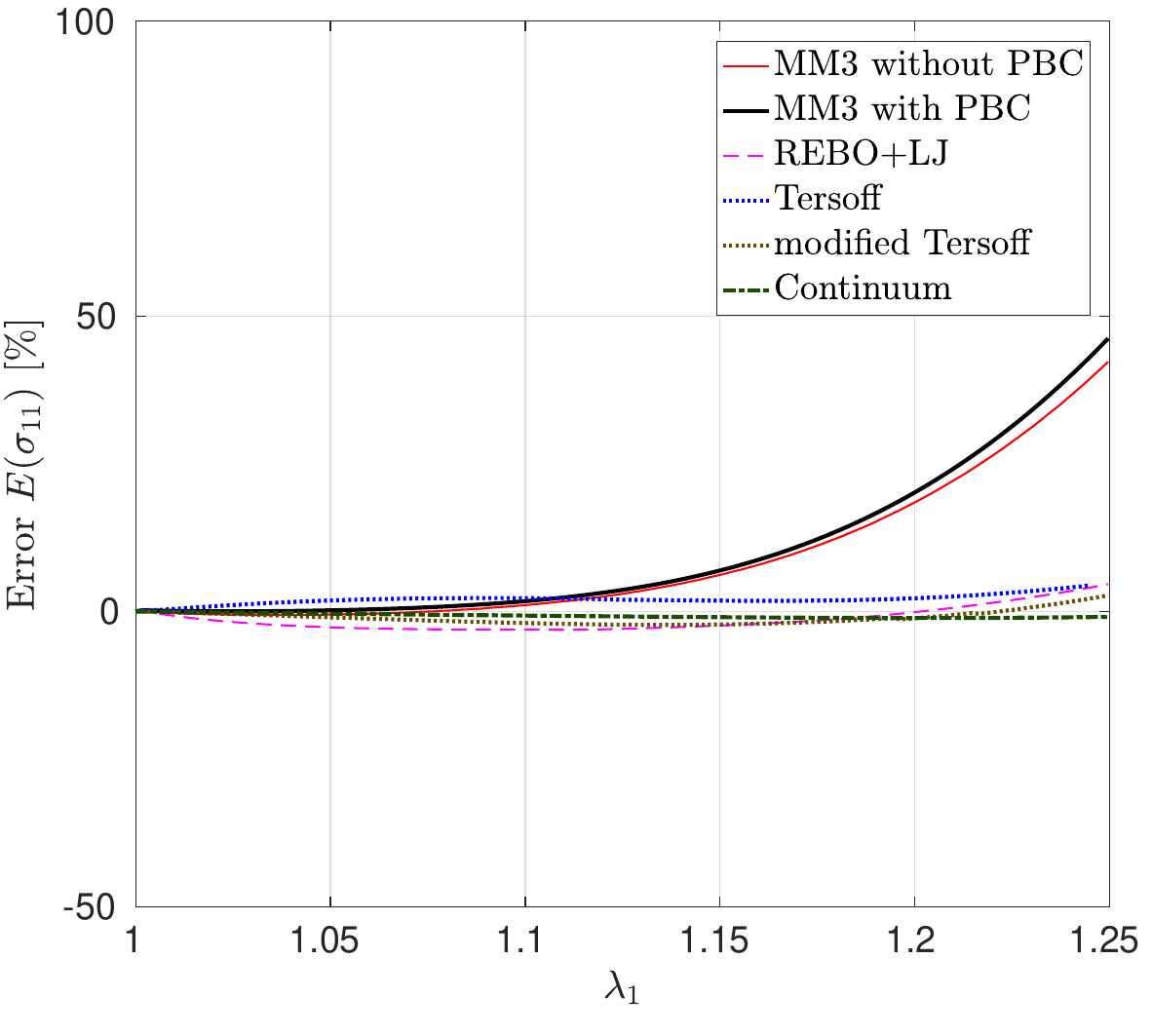}
        \vspace{-3mm}
        \subcaption{}
    \end{subfigure}
    \vspace{-3mm}
     \caption{Variation of (a) stress $\sigma_{11}$ and (b) error $E$($\sigma_{11}$) according to Eq.~(\ref{e:er_c}) as a function of stretch $\lambda_1$ in the zigzag direction.\label{fig:long_stress_ZZ}}
\end{figure}

\begin{figure}[!htbp]
        \begin{subfigure}[t]{0.49\linewidth}
        \centering
 \includegraphics[height=65mm]{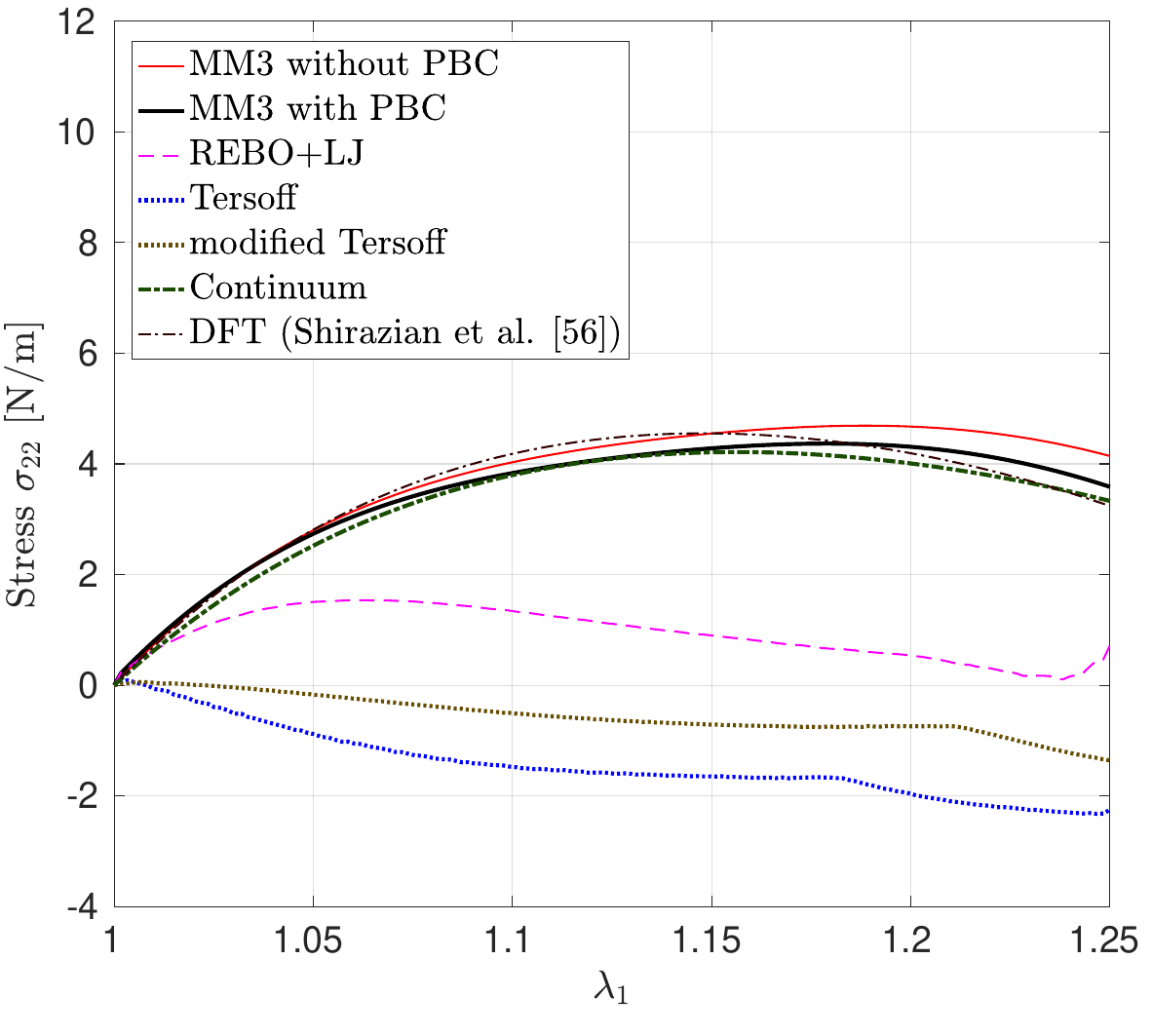}
         \vspace{-3mm}
        \subcaption{}
    \end{subfigure}
    \begin{subfigure}[t]{0.49\linewidth}
        \centering
    \includegraphics[height=65mm]{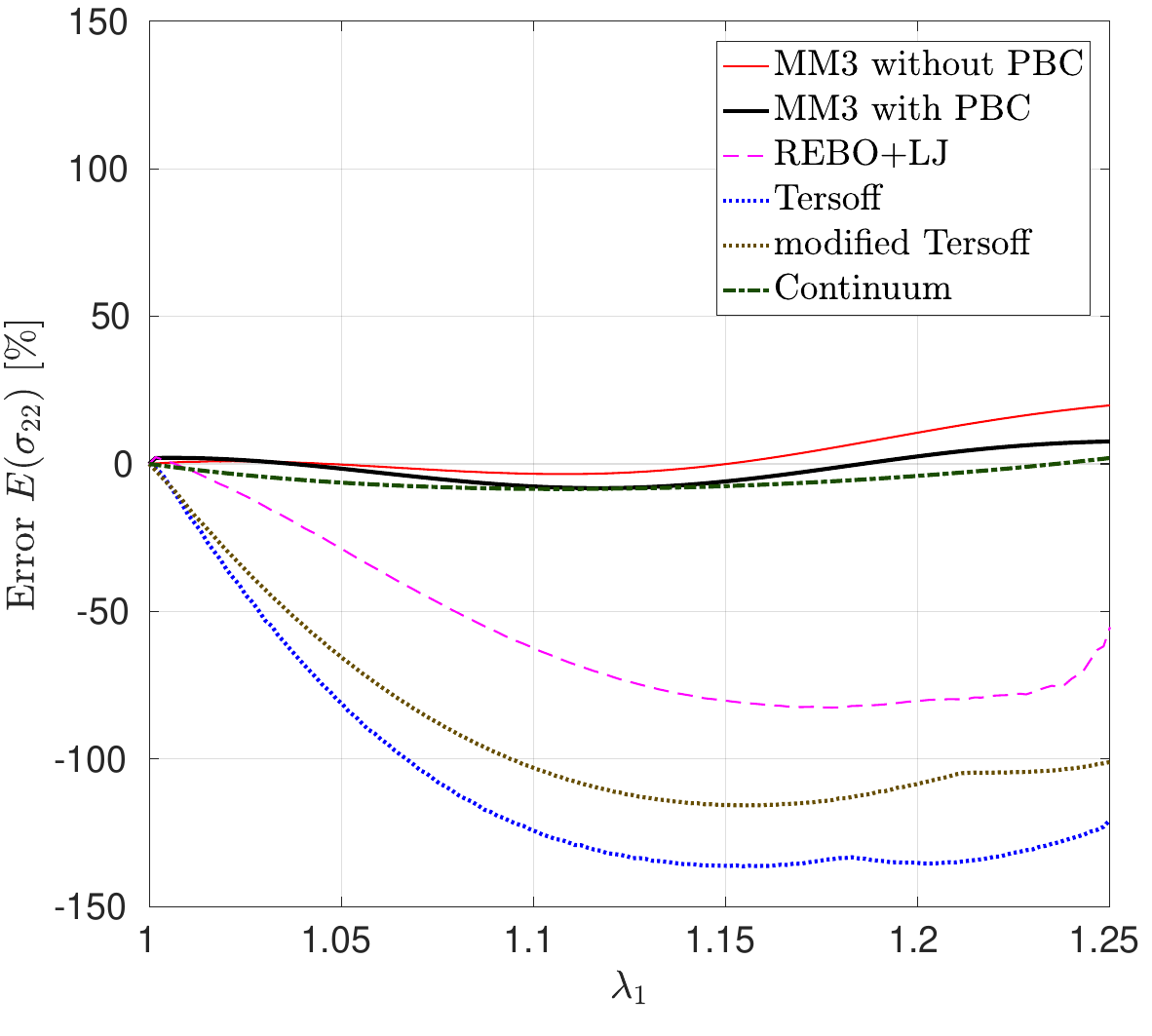}
        \vspace{-3mm}
        \subcaption{}
    \end{subfigure}
    \vspace{-3mm}
     \caption{Variation of (a) stress $\sigma_{22}$ and (b) error $E$($\sigma_{22}$) according to Eq.~(\ref{e:er_c}) as a function of stretch $\lambda_1$ in the armchair direction.\label{fig:lat_stress_AC}}
     \end{figure}

\begin{figure}[!htbp]
        \begin{subfigure}[t]{0.49\linewidth}
        \centering
 \includegraphics[height=65mm]{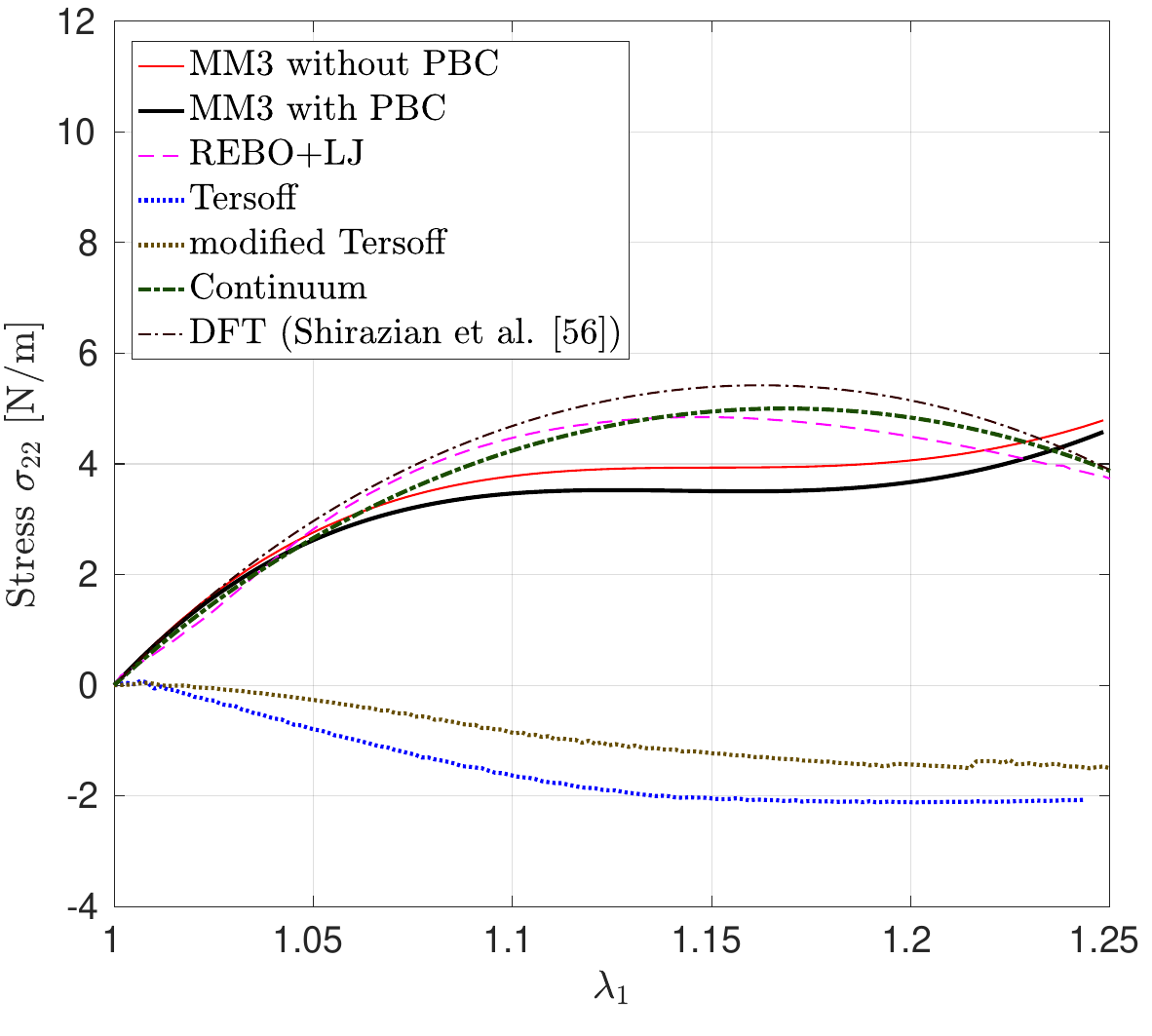}
         \vspace{-3mm}
        \subcaption{}
    \end{subfigure}
    \begin{subfigure}[t]{0.49\linewidth}
        \centering
    \includegraphics[height=65mm]{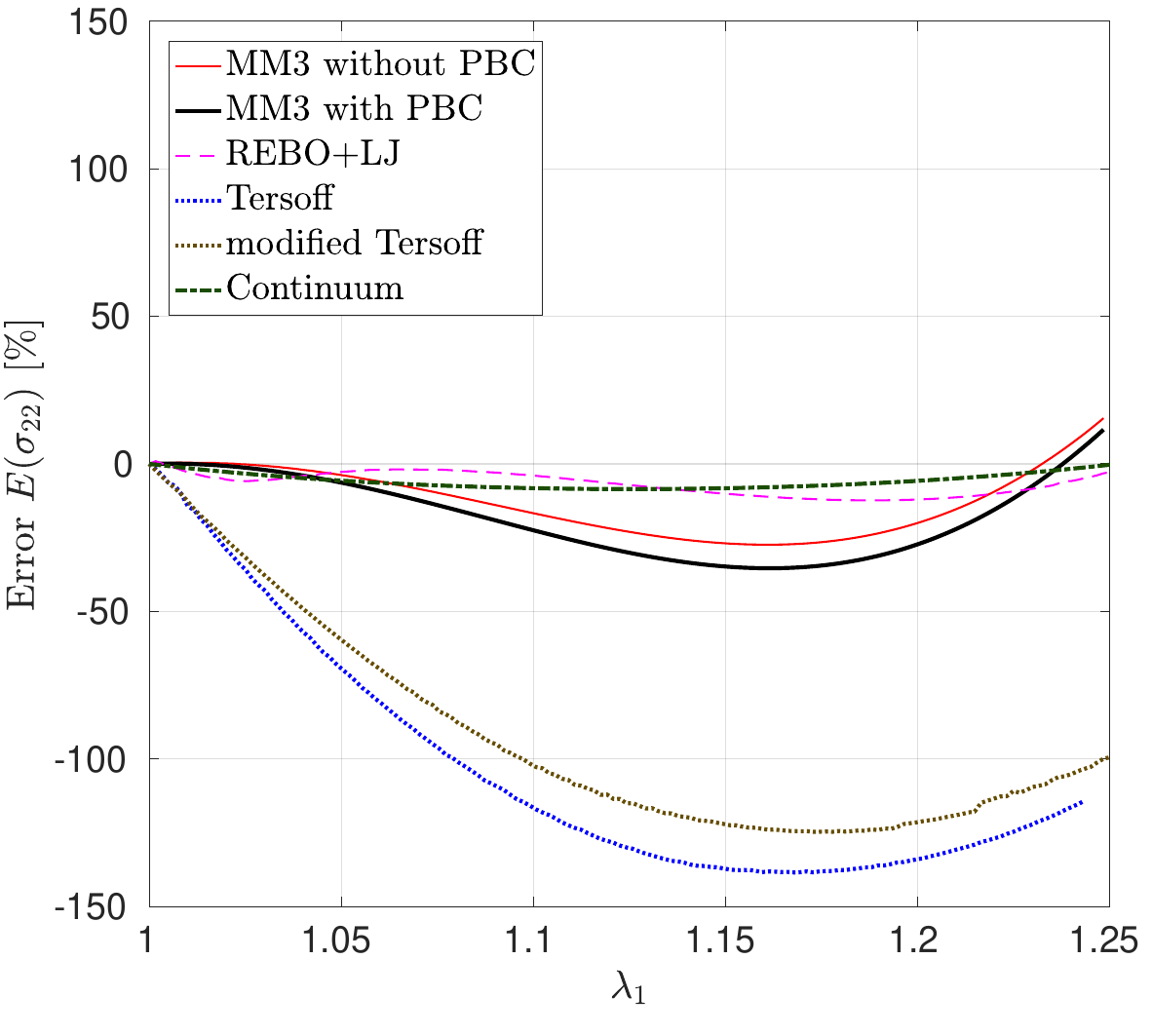}
        \vspace{-3mm}
        \subcaption{}
    \end{subfigure}
    \vspace{-3mm}
     \caption{Variation of (a) stress $\sigma_{22}$ and (b) error $E$($\sigma_{22}$) according to Eq.~(\ref{e:er_c}) as a function of stretch $\lambda_1$ in the zigzag direction.\label{fig:lat_stress_ZZ}}
\end{figure}

Next, the stresses from the different approaches are compared. For uniaxial stretch, the SLGS is stretched either along the armchair or zigzag direction. Three different stresses are examined in the following. \\
1. \textit{Stress along the stretch direction.}\\
The variation in $\sigma_{11}$ -- the stress along the stretch direction -- and the corresponding error are shown in Fig.~\ref{fig:long_stress_AC}
for stretch along the armchair direction, and Fig.~\ref{fig:long_stress_ZZ}, for stretch along the zigzag direction. Fig.~\ref{fig:long_stress_AC} shows that the molecular simulation results are in good agreement with the DFT results up to $\lambda_1 = 1.1 $ within an error of $\approx 5\%$.
Beyond this stretch, $\sigma_{11}$ computed from the MM3 potential shows  gradual hardening. $\sigma_{11}$ computed from the other two potentials follow the DFT results up to $\lambda_1=1.19$ and then there is a sudden rise, which is discusssed subsequently.
\\
Similarly Fig.~\ref{fig:long_stress_ZZ} shows that the molecular simulation results follow the DFT results up to $\lambda_1=1.13$.
Beyond this stretch, the MM3 results exhibit much higher stresses than the other cases. This may be due to the absence of cutoff function in MM3 as is present in the other two potentials, which initiates bond breaking in the structure. The results from the other two potentials continue to follow the DFT results up to $\lambda_1 = 1.25$.
\begin{figure}[!htbp]
    \begin{subfigure}[t]{0.49\linewidth}
        \centering
 \includegraphics[height=65mm]{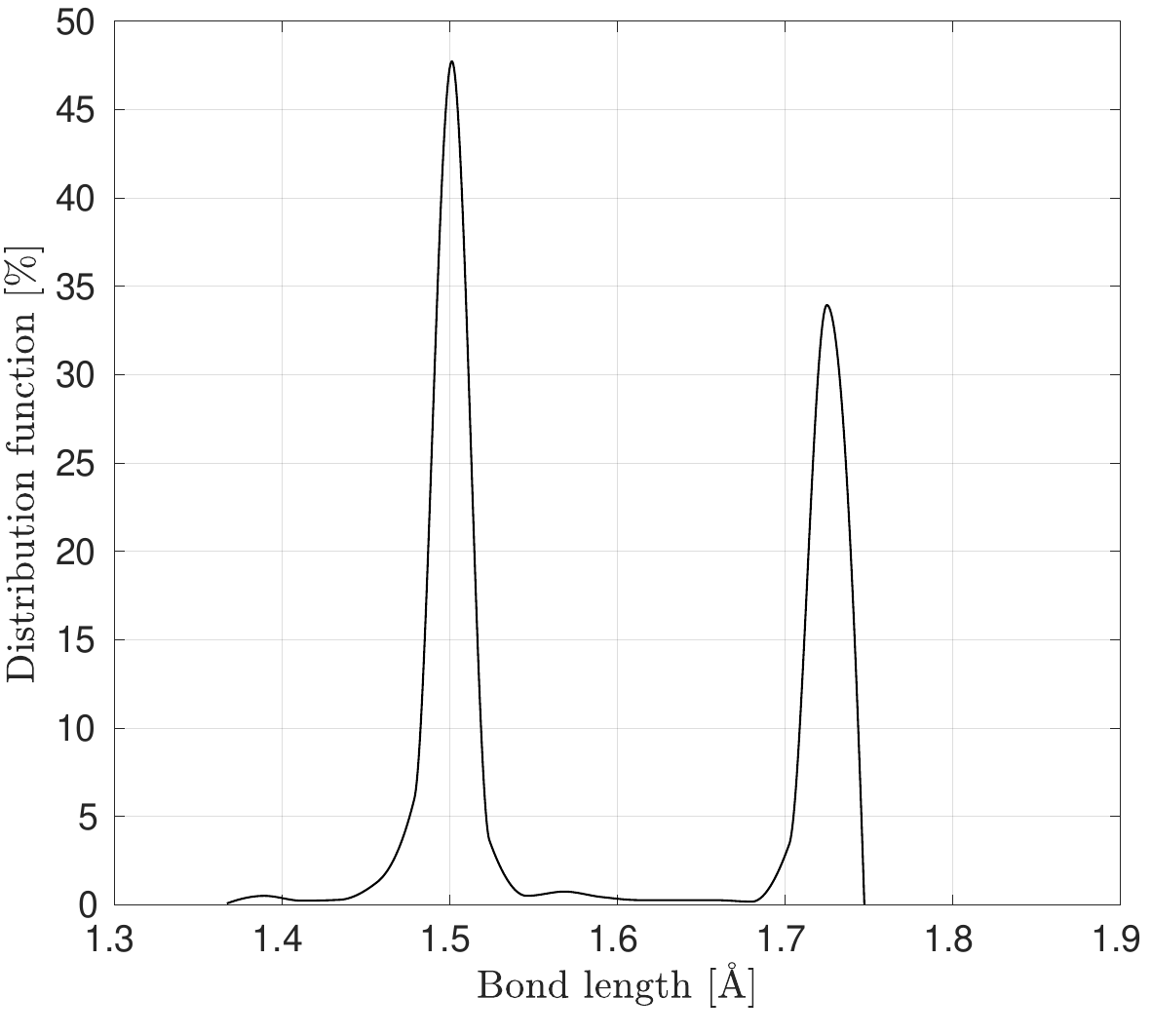}
         \vspace{-3mm}
        \subcaption{}
    \end{subfigure}
    \begin{subfigure}[t]{0.49\linewidth}
        \centering
    \includegraphics[height=65mm]{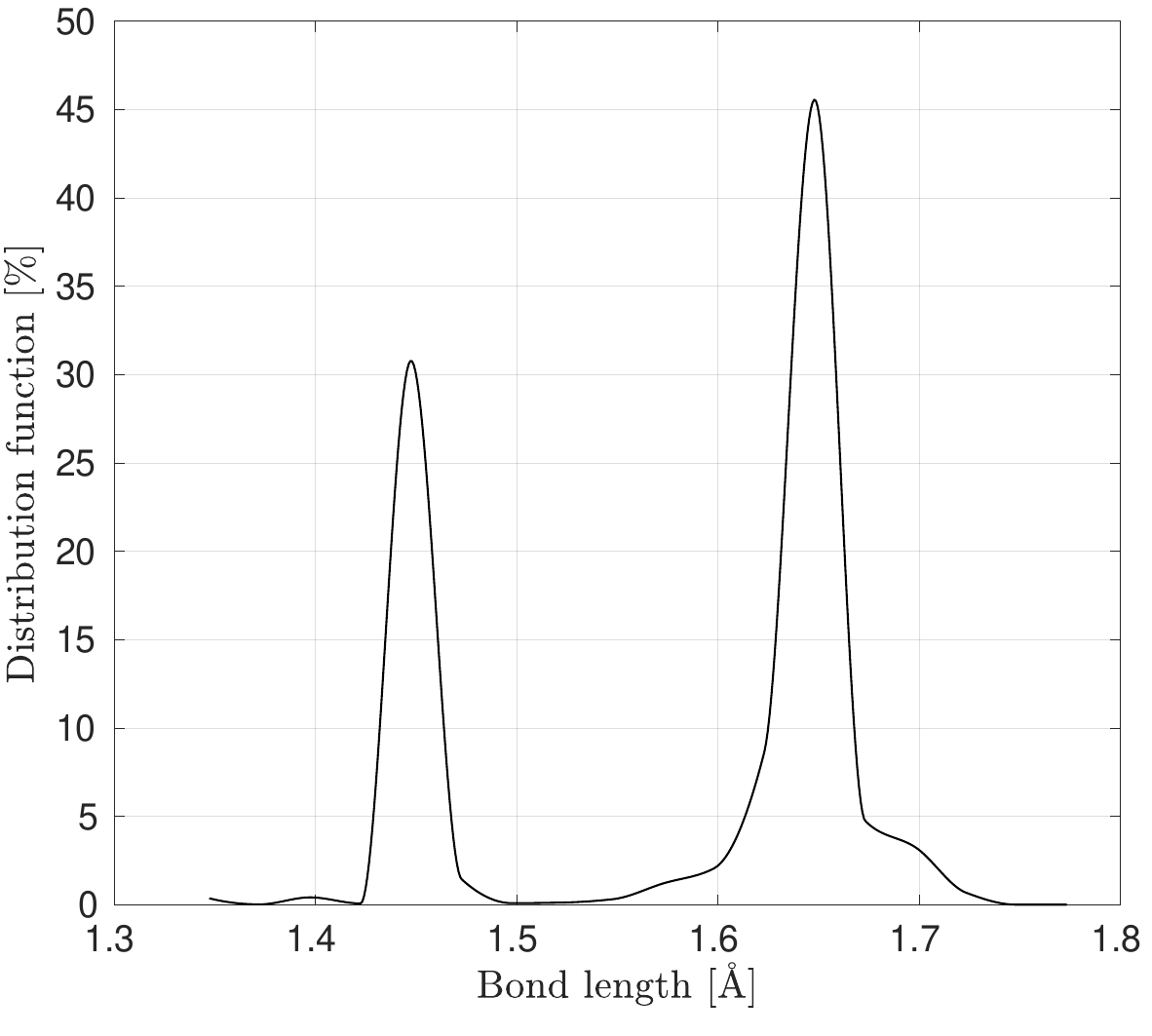}
        \vspace{-3mm}
        \subcaption{}
    \end{subfigure}
    \vspace{-3mm}
    \caption{Distribution function of the bond length for stretch along the (a) armchair and (b) zigzag directions at the stretch $\lambda_1=1.25$ for the REBO+LJ potential. The mean (and standard deviation) in nm of the two peaks for (a) and (b) are 0.1502 (0.0096), 0.1725 (0.00136) and 0.1447 (0.0058) and 0.1648 (0.0064), respectively. \label{fig:rebo_ac1}}
\end{figure} \\
As Fig.~\ref{fig:long_stress_AC} shows a sudden rise in $\sigma_{11}$ is noticed in the results with the REBO+LJ and Tersoff potentials in the armchair direction. In order to explain this behavior, Figs.~\ref{fig:rebo_ac1}a and \ref{fig:rebo_ac1}b show the distribution of the bond lengths at $\lambda_1 = 1.25$ for stretch along the armchair and zigzag directions, respectively, employing the REBO+LJ potential. It is found that around 36$ \% $ of bonds are stretched to a bond length of more than 0.17 nm when the stretch is along the armchair direction, which activates the cutoff function of Eq.~\eqref{eq:re_cut}. Due to the discontinuity of the second derivative of the cutoff function a sudden rise in the stress is recorded\footnote{REBO+LJ is sensitive to the choice of cutoff distances. Other values have shown to lead to bond breaking (and hence a sudden stress drop) at much lower strains.}. In the case of the Tersoff potential, the anomalous response in $\sigma_{11}$ in the armchair direction around $\lambda_1 = 1.2$ is due to the elongation of bonds beyond the cutoff radii 0.18 and 0.185 nm for the Tersoff and modified Tersoff potentials, respectively, see Eq.~\eqref{tersoff_cut}. \\
2. \textit{Stress perpendicular to the stretch direction.}\\
The variation in the lateral stress $\sigma_{22}$ -- the stress in the perpendicular direction to the stretch -- and its error according to Eq.~\eqref{e:er_c} as a function of $\lambda_1$ are shown in Figs.~\ref{fig:lat_stress_AC} and \ref{fig:lat_stress_ZZ}. The figures show that the results from the MM3 potential agree well with the DFT results up to $\lambda_1 =  1.15$ and $\lambda_1 = 1.1$ in the armchair and zigzag directions, respectively. As Fig.~\ref{fig:lat_stress_AC} shows for stretch in the armchair direction, the lateral stresses from the REBO+LJ potential agree well with the DFT results within the small deformation regime (upto $\lambda_1=1.025$) and then show gradual softening. However, for stretch in the zigzag direction (see Fig.~\ref{fig:lat_stress_ZZ}), $\sigma_{22}$ from REBO+LJ matches well with $\sigma_{22}$ from DFT up to $\lambda_1 =  1.25$. Contrary to the results from the MM3 and REBO+LJ potentials, both Tersoff potentials produce negative lateral stress for uniaxial stretch. \\
     
\begin{figure}[!htbp]
    \begin{subfigure}[t]{1\linewidth}
        \centering
     \includegraphics[height=65mm]{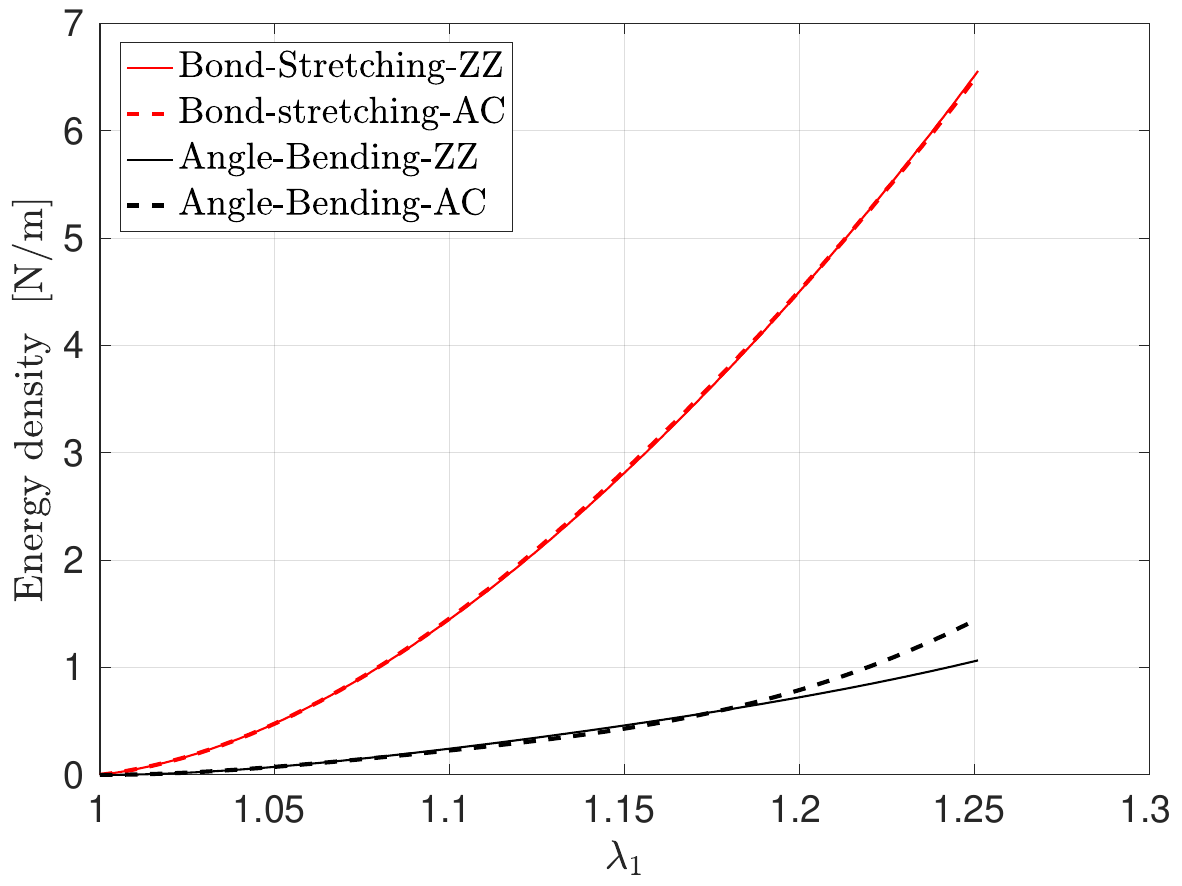}
         \vspace{-3mm}
        \end{subfigure}
    \caption{Comparison of the dominant energy terms of the MM3 potential when the square SLGS is stretched along the armchair (AC) and zigzag directions (ZZ). \label{fig:mm3_ene_parts}}
\end{figure}

\begin{figure}[!htbp]
        \begin{subfigure}[t]{0.49\linewidth}
        \centering
 \includegraphics[height=65mm]{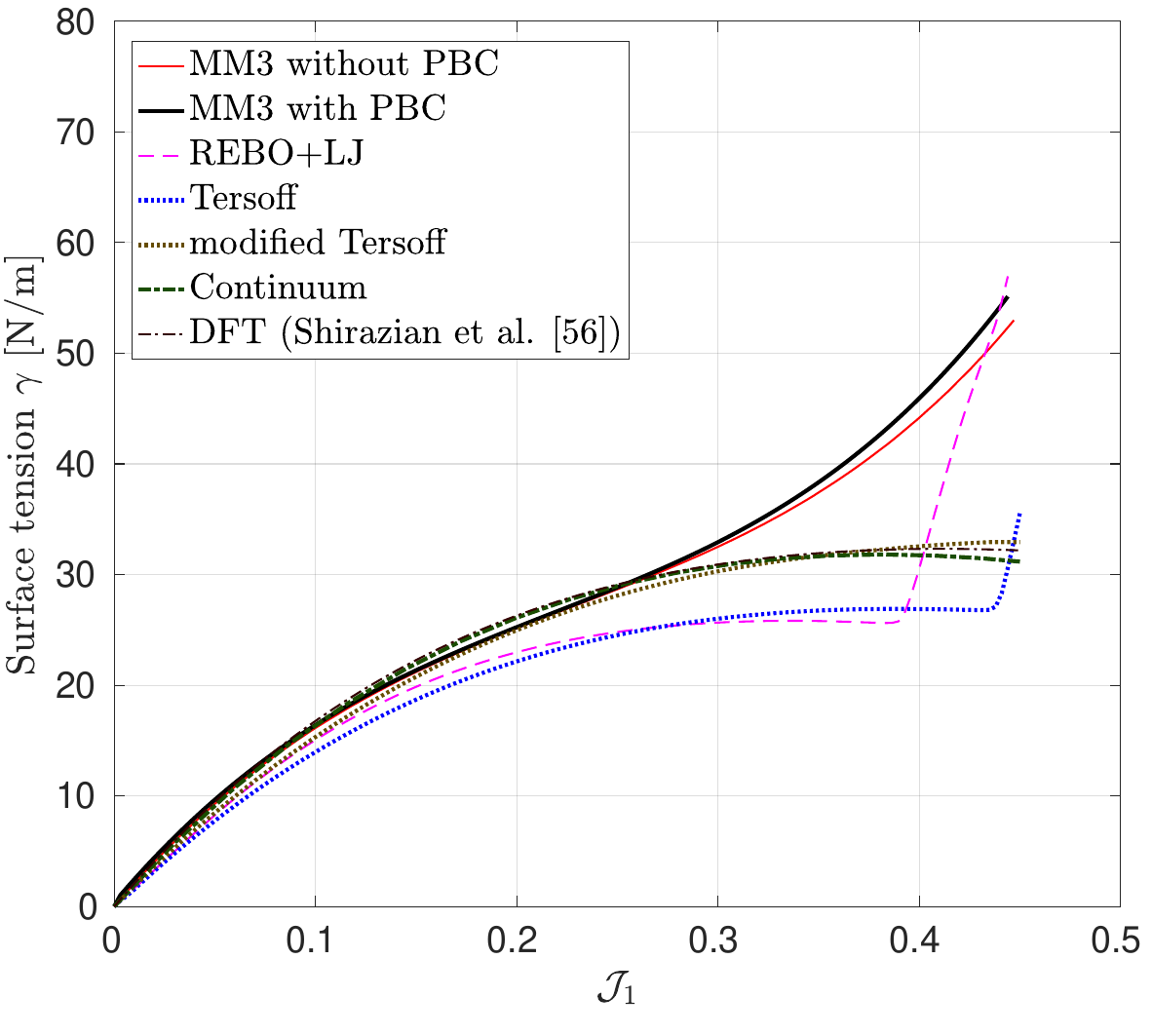}
         \vspace{-3mm}
        \subcaption{}
    \end{subfigure}
    \begin{subfigure}[t]{0.49\linewidth}
        \centering
    \includegraphics[height=65mm]{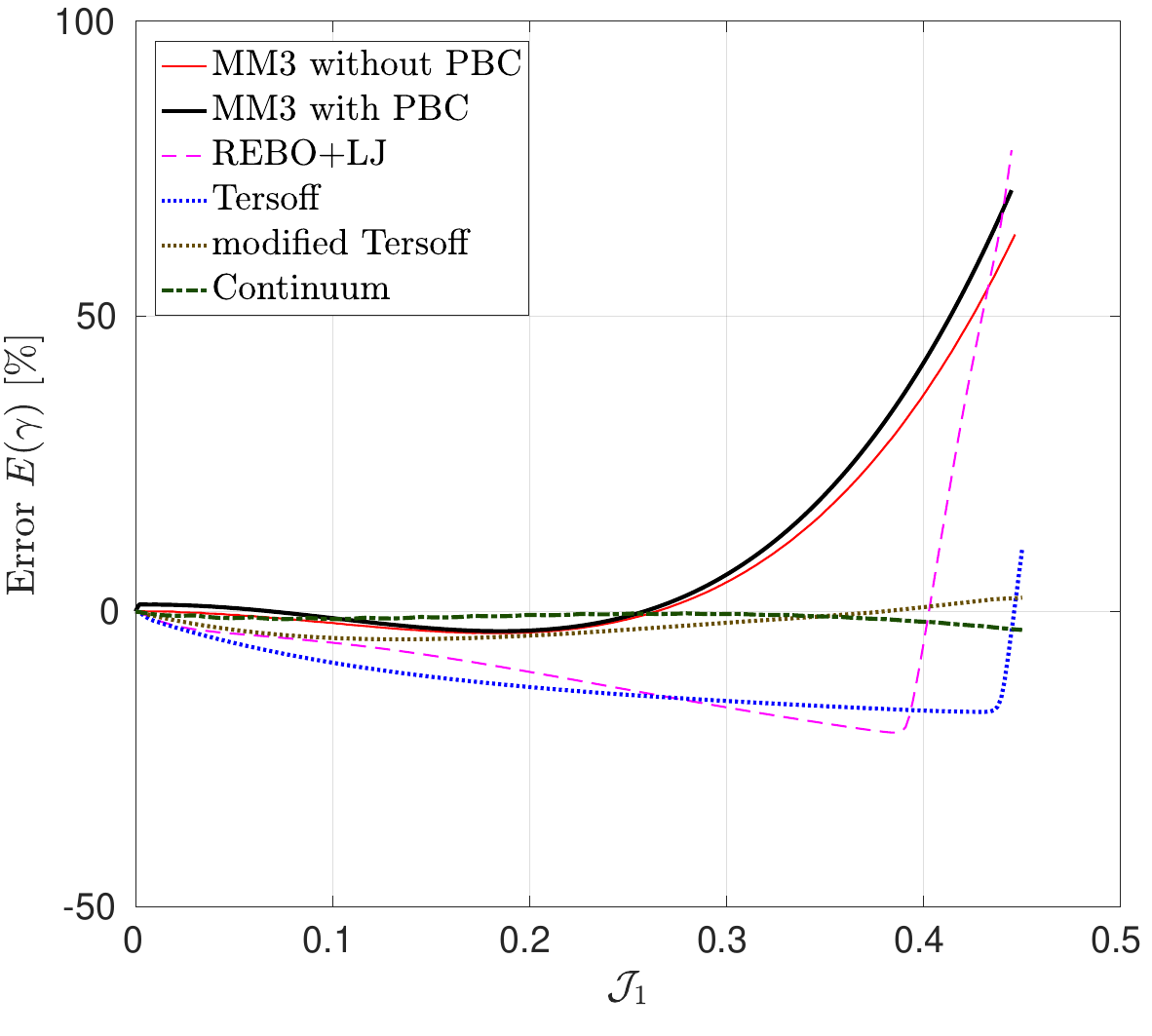}
        \vspace{-3mm}
        \subcaption{}
    \end{subfigure}
    \vspace{-3mm}
     \caption{Variation of (a) surface tension $\gamma$ and (b) error $E(\gamma)$ according to Eq.~(\ref{e:er_c}) as a function of dilatation $ \mathcal{J}_1 $. \label{fig:dilatation}}
\end{figure}

To explain this exceptional behavior of the Tersoff potential, we have performed simulations without the constraints on the lateral edge atoms and calculated the Poisson's ratio $\nu = -{\varepsilon_{\text{lat}}}/{\varepsilon_{\text{lon}}}, $ where $ \varepsilon_{\text{lat}}$ and $\varepsilon_{\text{lon}}$ are the lateral and longitudinal strains, respectively. The strains are computed by taking the ratio of the change in periodic box dimensions with the initial box dimensions. For stretch along the zigzag and armchair directions, the SLGS exhibits a negative Poisson's ratio for all the stretch ratios according to the Tersoff potential. This has also been reported in the literature \citep{Berinskii2010_01,Sgouros2016_01,Lebedeva2019}. Due to this behavior, the Tersoff potential stands in sharp contrast to all the other methods.
      \\
It is noticed that the stress response computed from the continuum model for both the loading cases are in good agreement with those from DFT. Both the molecular and continuum stresses are equal at smaller stretches, and it is evident that the SLGS shows anisotropy at higher stretches. Additionally, the anisotropy predicted from both the approaches is different, i.e., molecular simulations predict that the armchair direction is stiffer whereas DFT and continuum predict otherwise. \\
To elucidate the anisotropy of the SLGS, we have plotted the two dominating energy terms\footnote{Van der Waals interactions play an insignificant role here. Including them reveals that they are below  0.06 N/m, which is less than 1$\%$ of the bond-stretching term.} of the MM3 potential in Fig.~\ref{fig:mm3_ene_parts}. The figure shows that the energy contribution from bond-stretching (term $U_s$ in Eq.~(\ref{eq:mm3})) is the same for stretch in both directions.
The contribution of angle-bending on the other hand (term $U_\theta$) to the total energy for both cases is equal up to $\lambda_1 = 1.17$, but deviates beyond this stretch. It increases more strongly when the SLGS is stretched in the armchair direction. The remaining energy contributions in Eq.~(\ref{eq:mm3}) are not significant for this deformation state. Thus, we conclude that $U_\theta$ is responsible for the anisotory in SLGS at large deformations. \\
3. \textit{Surface tension.}\\
Finally, the variation in the surface tension $\gamma = (\sigma_{11}+\sigma_{22})/2$, under pure dilatation and its corresponding percentage error is shown in Fig.~\ref{fig:dilatation}.
As seen, the results from molecular simulation employing the MM3 potential agree with the results from DFT up to $\mathcal{J}_1 = 0.28$ and then show gradual hardening, whereas the results from the REBO+LJ and Tersoff potentials agree only up to $ \mathcal{J}_1 = 0.08$ and then deviate. Additionally, the results from the later two potentials exhibit a sharp rise at $\mathcal{J}_1 = 0.39$ and 0.43, respectively. At those values of $\mathcal{J}_1$,  $\approx 96 \%$ of bonds are stretched to more than 0.17 nm and 0.18 nm ($r_{IJ}^{\mathrm{min}}$ of respective potentials) using the REBO+LJ and Tersoff potentials, respectively. On the other hand, the results from the continuum model and the modified Tersoff potential agree within an error of $\approx 5 \%$.
\subsection{Out-of-plane bending of SLGS} \label{out_ben}
\begin{figure}[!htbp]
    \begin{subfigure}[t]{0.49\linewidth}
        \centering
 \includegraphics[height=65mm]{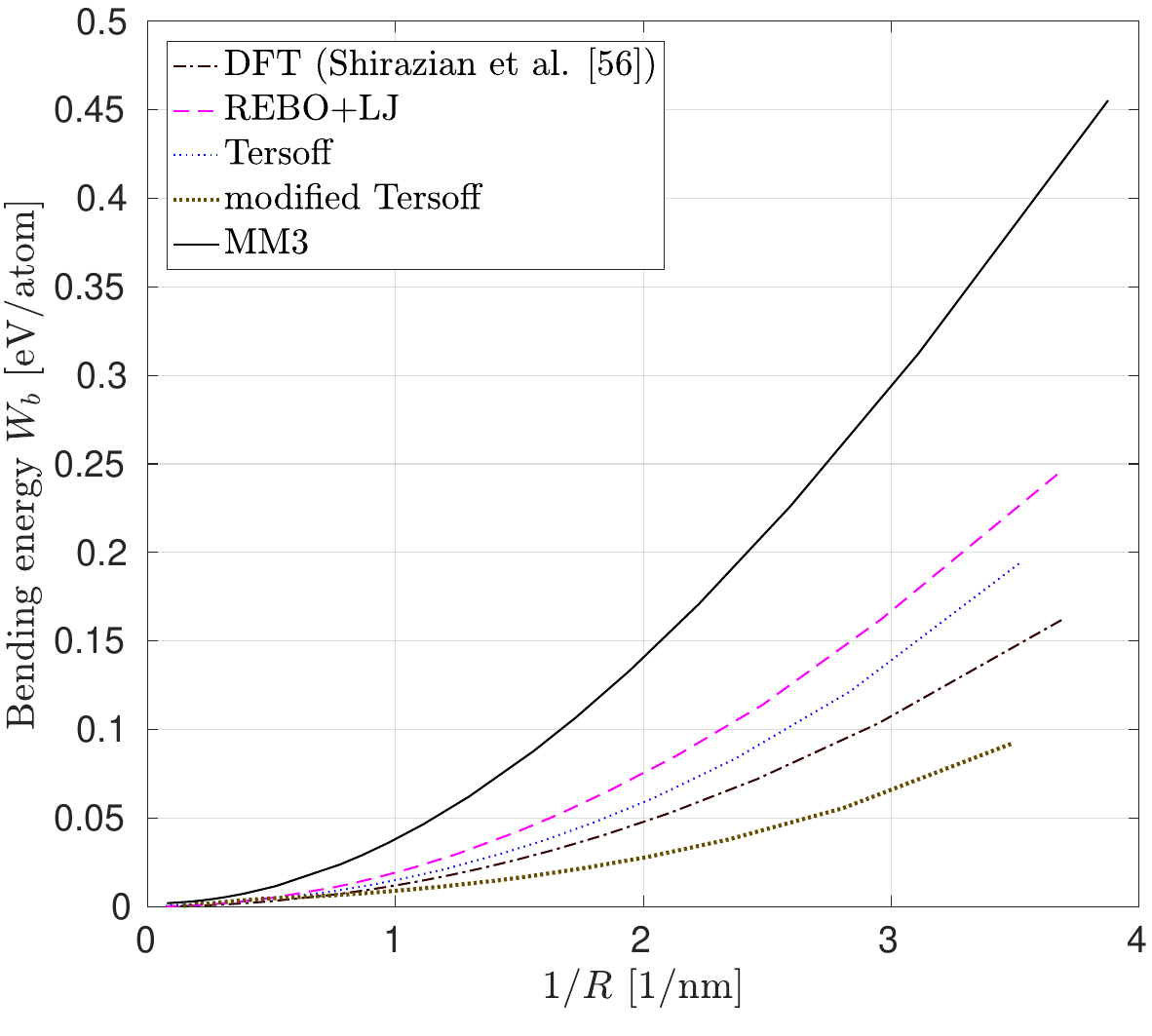}
         \vspace{-1mm}
        \subcaption{}
    \end{subfigure}
    \begin{subfigure}[t]{0.49\linewidth}
        \centering
    \includegraphics[height=65mm]{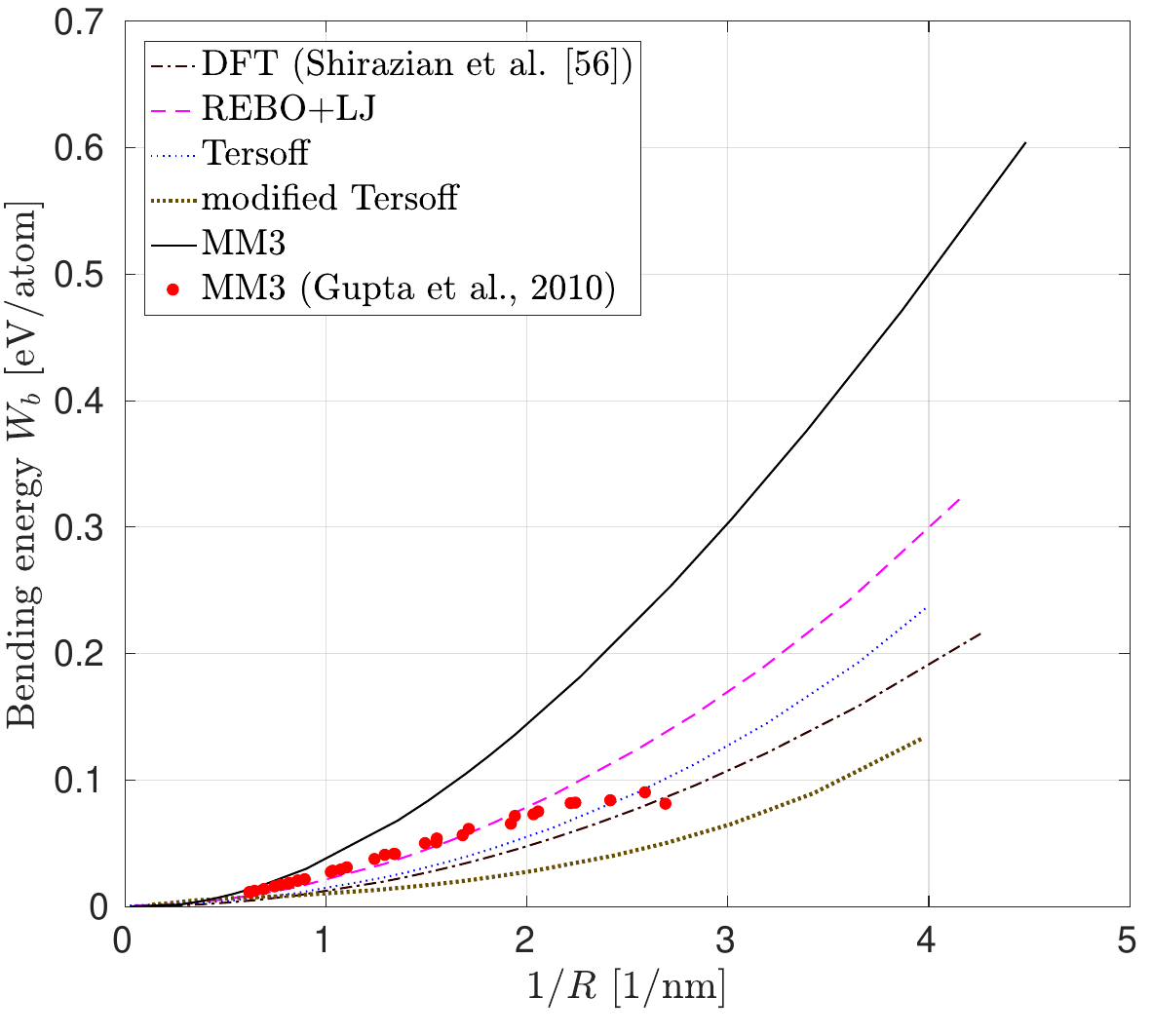}
        \vspace{-1mm}
        \subcaption{}
    \end{subfigure}
    \vspace{-3mm}
    \caption{Variation of the bending energy with the bending curvature for (a) armchair and (b) zigzag CNTs.\label{fig:rebo_ac}}
\end{figure}

In molecular simulations, the bending stiffness of the SLGS can be calculated in two different ways. They both assume linear elastic bending behavior. \\
In the first approach, the SLGS is considered as a thin, linearly elastic, homogeneous and isotropic plate with thickness $h$, whose bending stiffness is given by \citep{Timoshenko1959a}
\eqb{lll} \label{eq:bend}
 c_{\text{b}} =\ds \frac{Eh^3}{12(1-\nu^2)}~,
\eqe
where $E$ and $\nu$ are the Young's modulus and Poisson's ratio, respectively. To avoid introducing a thickness of the SLGS in the first approach, the first bending frequency of a square SLGS is computed instead in our work. In MS simulations, the \textit{Vibrate} module in Tinker is used to find the natural  frequencies. In MD, on the other hand, the equilibrated SLGS  is deformed into the mode shape corresponding to the first natural frequency and then allowed to vibrate freely in the NVE ensemble. The time history of the atoms closest to the center of the SLGS is then  evaluated to determine the frequencies of the transverse vibration using fast Fourier transform (FFT). The obtained frequencies are then compared with those determined from an equivalent plate model. The formula for the transverse frequencies of a linearly elastic, homogeneous and isotropic square plate of side $a$ is given by  \citep{blevins_1979}
\eqb{lll} \label{eq:bend1}
f_{mn} =\ds \frac{\lambda_{mn}^2}{2\pi a^2}\sqrt{ \frac{c_\mathrm{b}}{\rho_\mrs}}~,
\eqe
where $m$ and $n$ are the half wave numbers along the x and y directions, $c_\mathrm{b}$ and $\rho_\mrs$ are the bending stiffness and areal density, respectively. For $m=1$ and $n=1$, the constant $\lambda_{11}^2$ is 35.99 \citep{blevins_1979}.
\begin{table}[!htbp]
\footnotesize
\centering
\caption{Bending stiffness ($c_{\text{b}}$) obtained from an equivalent plate model. Here $K$, $\nu$ and $h$ are the basal plane stiffness, Poisson ratio and the effective thickness of a graphene sheet, respectively. Here, first three rows of the table are calculated from Eq.~\ref{eq:bend} and remaining are from Eq.~\ref{eq:bend1}. \label{bending}}
\begin{tabular}{lcccc}
  \hline
      Potential/model & $K$ [N/m] &  $\nu$ & $h$ [nm]  & $c_{\text{b}}$ [eV] \\
    \hline
MM3 \citep{Gupta2010a} & 340.0 & 0.210 & 0.100 & 1.86 \\
REBO-I \citep{Huang2006} & 235.0 & 0.412 & 0.062 & 0.56 \\
REBO-II \citep{Huang2006} & 243.0 & 0.397 & 0.057 & 0.49  \\
MM3 (Current work) & - & - & - & 2.11  \\
REBO+LJ (Current work) & - & - & - & 2.27   \\
Tersoff (Current work)& - & - & - & 2.08  \\
modified Tersoff (Current work)& - & - & - & 0.72 \\
  \hline
\end{tabular}
\end{table}
The bending stiffness determined from the MM3, REBO+LJ and Tersoff potentials, according to Eqs.~(\ref{eq:bend}) and (\ref{eq:bend1}) (see Table \ref{bending}) is larger than 1.49 eV, which is the value from DFT \citep{Shirazian2018_01, Kudin2001_01}. The discrepancy may be associated with the presence of pre-stress in the molecular simulations, which also causes higher frequencies at zero stretch. This is discussed further below. On the other hand, the bending stiffness determined from the modified Tersoff potential is 51 $\%$ lower than the DFT value.\\
In the second approach, the bending energy of the SLGS is obtained by computing the potential energy of relaxed carbon nanotubes of different radii with respect to the ground state energy of SLGS. The potential energy is fitted by a quadratic curve. The second derivative of this curve then corresponds to the bending stiffness.  The obtained values are listed in Table~\ref{bending_non}. We note that the second approach is problematic, as relaxed CNTs usually change radius and therefore also contain in-plane strain energy that is usually not accounted for in the stiffness calculation. As a consequence the bending stiffness may be overestimated.
\begin{table}[h]
\footnotesize
\centering
\caption{Bending stiffness $c_{\text{b}}$ in [eV] from an equivalent CNT model. \label{bending_non}}
\begin{tabular}{lcccc}
  \hline
  Potential/model & $c_{\text{b}}$ (armchair) & $c_{\text{b}}$ (zigzag) \\
    \hline
   MM3 (Current work)   & 3.271 & 3.146\\
   REBO+LJ (Current work)   & 2.184 & 2.235 \\
   Tersoff (Current work) & 2.078 &  2.010 \\
   modified Tersoff (Current work) & 0.915 &  0.969 \\
  \hline
\end{tabular}
\end{table} \\
The variation of the bending energy with the curvature for armchair and zigzag carbon nanotubes is shown in Fig.~\ref{fig:rebo_ac}a and \ref{fig:rebo_ac}b. The bending energy from the DFT and DFT-equivalent continuum model in (\ref{eq:b_c}) is given by
$W_{\text{b}}=\ds {c_{\text{b}}}\kappa_1^2/{2}~,$ where $c_{\text{b}}$ is the bending stiffness and $\kappa_1$ is the curvature radius. The curvature radius of a CNT is the reciprocal of the radius $ R =({\sqrt3a_{cc}}/{2\pi})\sqrt{n^2+m^2+nm}$~, where $a_{cc}$ is the equilibrium bond length of C-C, and $n$ and $m$ are the chirality indices.
The bending energy calculated from the MM3 potential differs more than the other two potentials considered. The difference may be attributed to the presence of higher order and cross-interaction terms of the potential. In their study, \citet{Gupta2010} computed the bending stiffness of different radii and chiralities of CNTs by equating the frequencies from MS simulations employing the MM3 potential with that of a shell model. The bending energy computed from these calculations is shown in Fig.~\ref{fig:rebo_ac}(b).
\subsection{Modal analysis} \label{freq}
In the following we discuss the effect of incremental prestretch in a SLGS and a CNC on their first few modes of vibrations.
\subsubsection{Square SLGS}
After having established the validity and accuracy of the MM3 potential for stretches up to $\approx 1.1$ in the previous section, we now study the transverse modes of vibrations of a SLGS using this potential. The vibration response is then compared with that obtained from the DFT-based continuum model at different stretch states. In the continuum model, the stretch is applied up to the loss of ellipticity of the elasticity tensor. This limiting value of stretch is also used in the molecular simulations.
\begin{figure}[h]
    \begin{subfigure}[t]{1\linewidth}
        \centering
     \includegraphics[width=70mm]{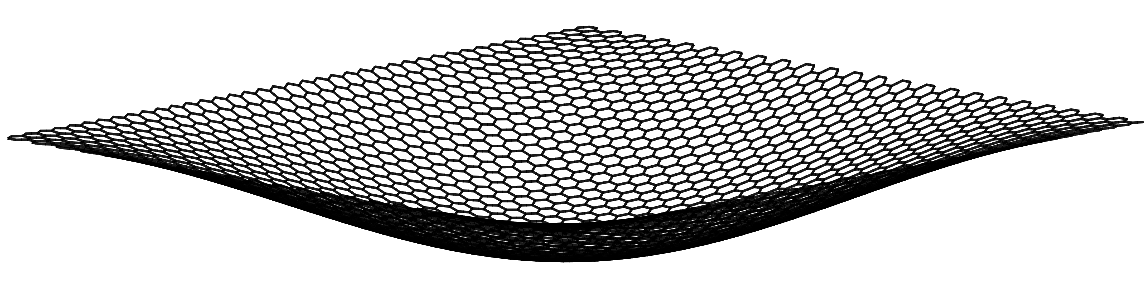}
         \vspace{-3mm}
        \subcaption{}
    \end{subfigure}
    \begin{subfigure}[t]{0.49\linewidth}
        \centering
 \includegraphics[width=70mm]{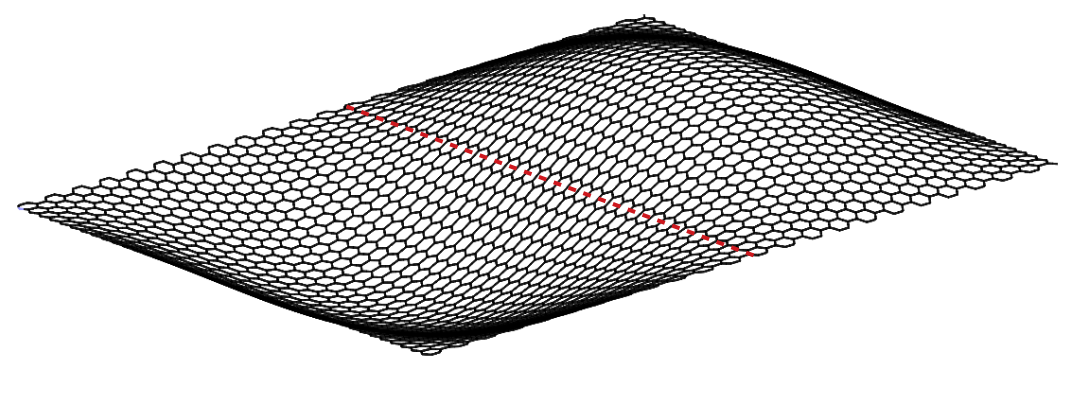}
         \vspace{-3mm}
        \subcaption{}
    \end{subfigure}
    \begin{subfigure}[t]{0.49\linewidth}
        \centering
    \includegraphics[width=70mm]{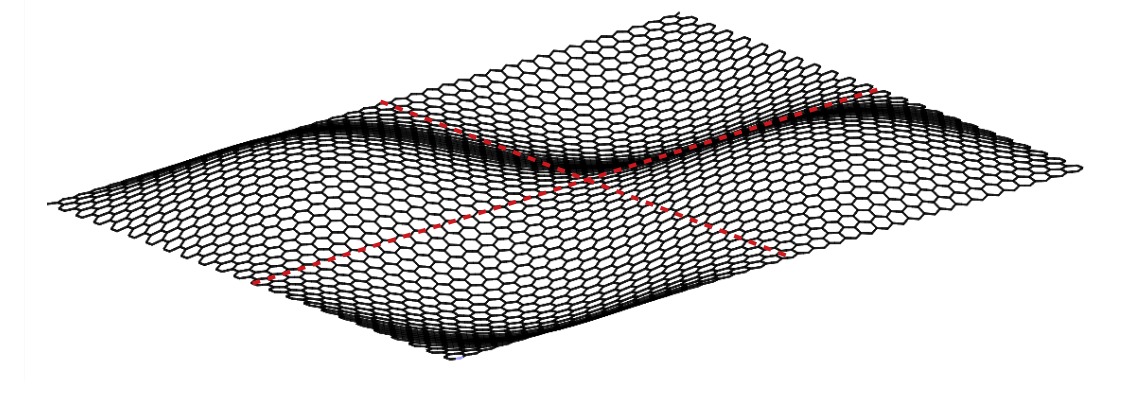}
        \vspace{-3mm}
        \subcaption{}
    \end{subfigure}
    \caption{Mode shapes of a relaxed square graphene sheet determined by a MS simulation: Shape corresponding to frequency (a) $ \omega_{11} $, (b) $ \omega_{12} $, and (c) $ \omega_{22} $. Here the two indices in $\omega$ represent the number of half waves along the local $x-$ and $y$-axis, respectively, and the dotted red lines show the nodal lines. \label{fig:modes}}
\end{figure}
\begin{figure}[!htbp]
    \begin{subfigure}[t]{0.49\linewidth}
        \centering
 \includegraphics[width=70mm]{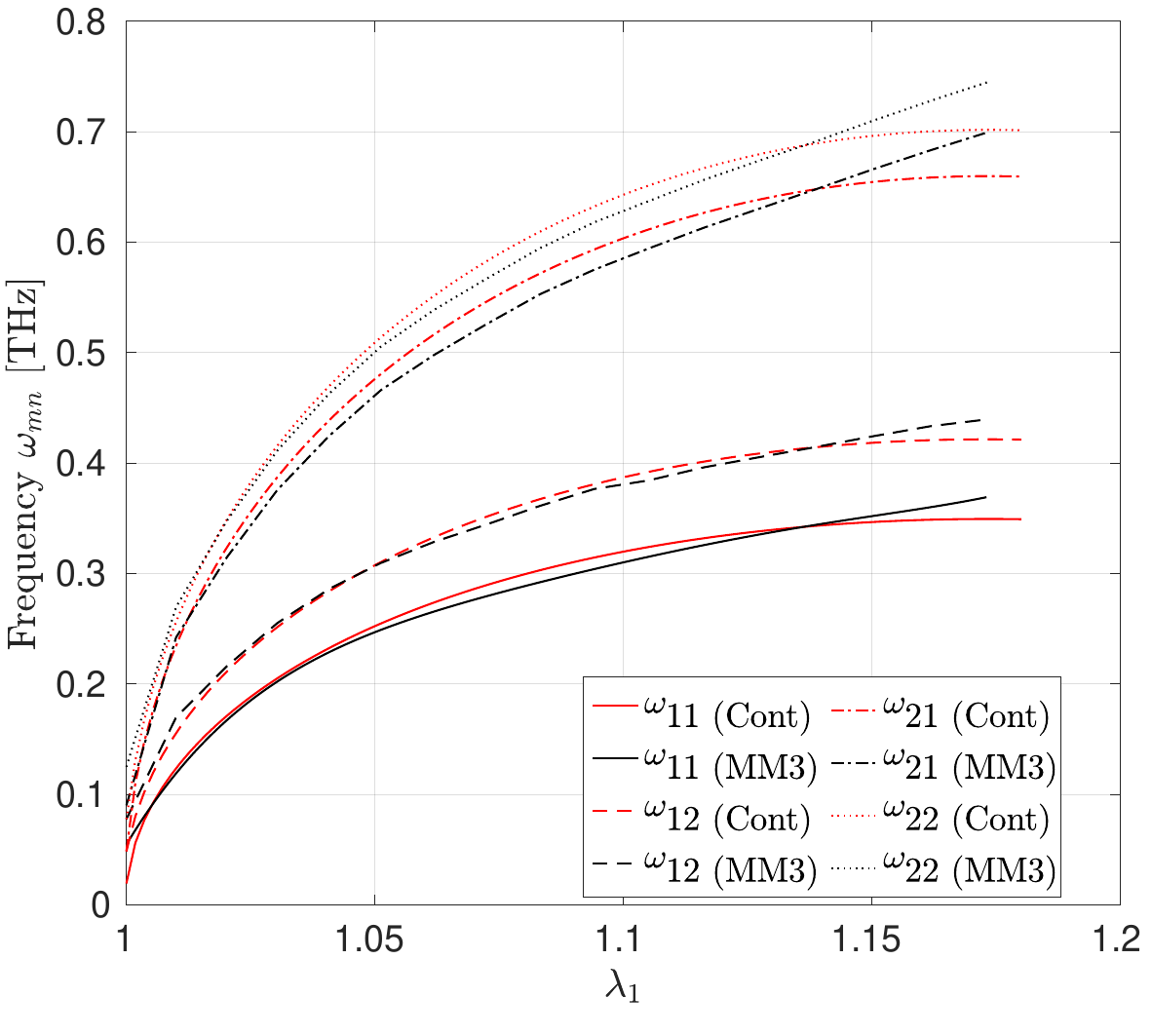}
         \vspace{-3mm}
        \subcaption{}
    \end{subfigure}
    \begin{subfigure}[t]{0.49\linewidth}
        \centering
    \includegraphics[width=70mm]{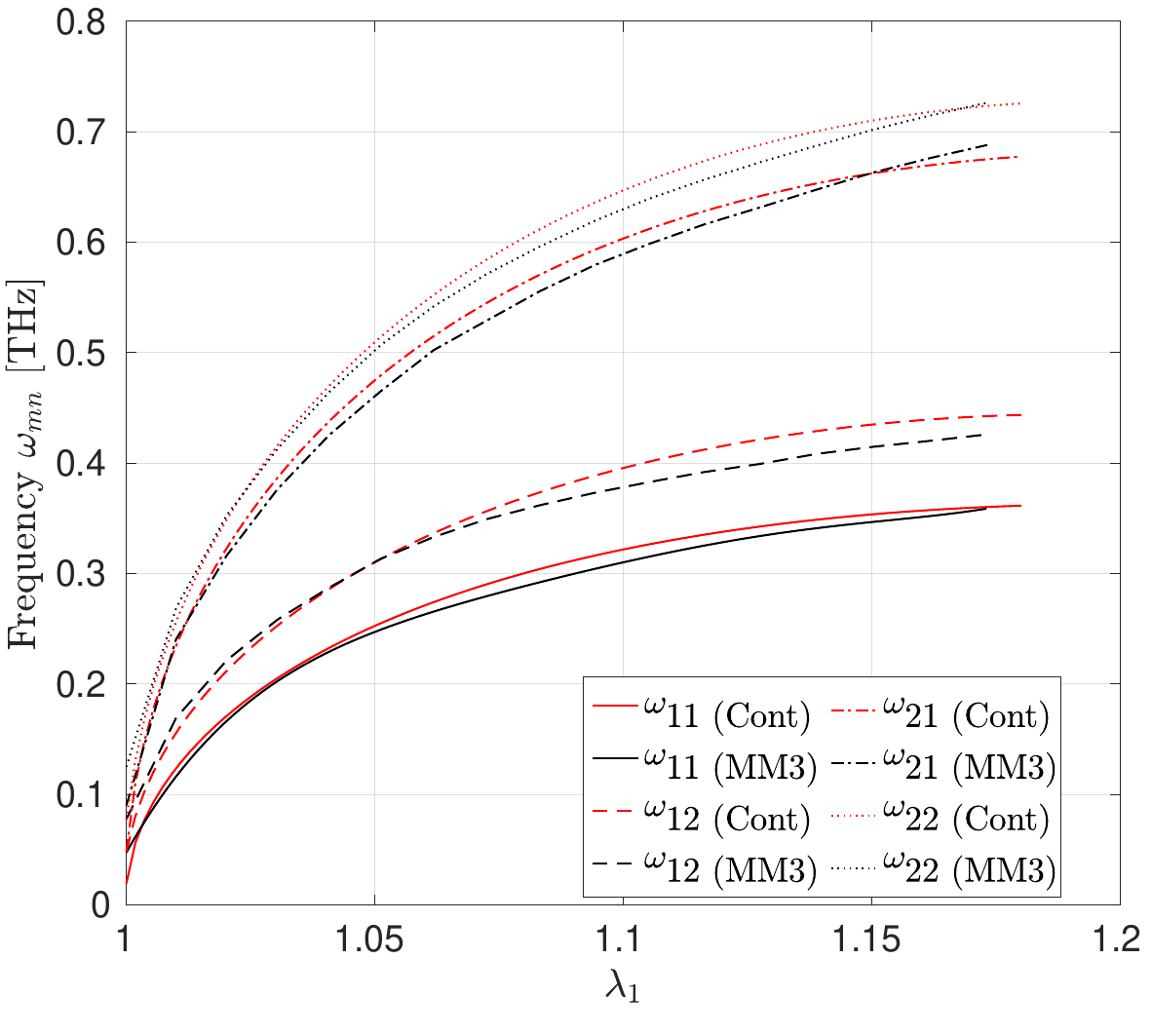}
        \vspace{-3mm}
        \subcaption{}
    \end{subfigure}
    \begin{subfigure}[t]{1\linewidth}
        \centering
     \includegraphics[width=70mm]{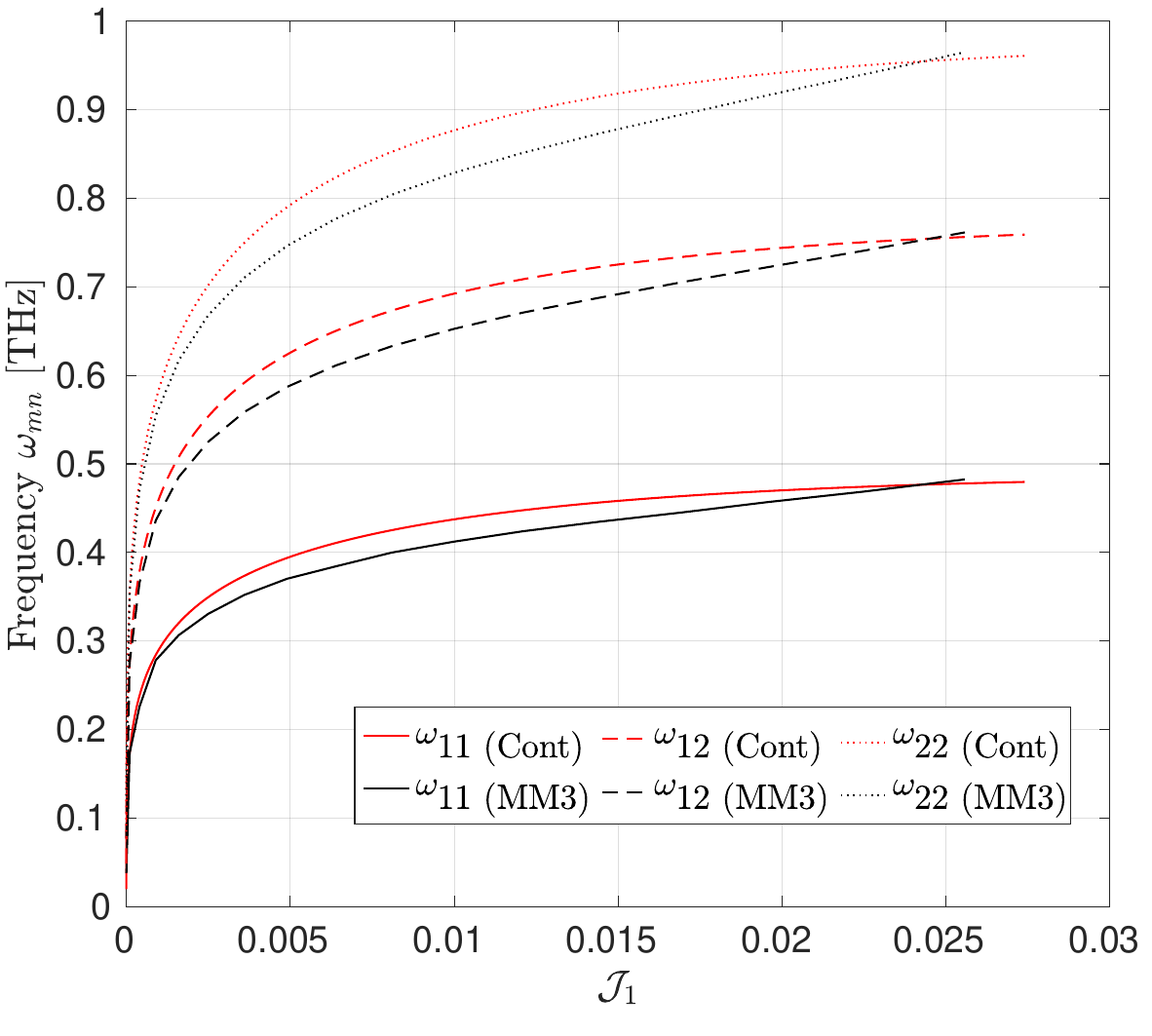}
         \vspace{-3mm}
        \subcaption{}
    \end{subfigure}
    \vspace{-3mm}
    \caption{Frequencies of a 10 nm $\times$ 10 nm, all-edge-clamped graphene sheet under (a) uniaxial stretch along the armchair direction, (b) uniaxial stretch along the zigzag direction and (c) pure dilatation. \label{fig:Gra_frequency_comparsion_L10_10}}
\end{figure}
The mode shapes and frequency variation of the transverse vibrations of the first three modes of the SLGS with increasing stretch/dilatation are shown in Fig.~\ref{fig:modes} and Fig.~\ref{fig:Gra_frequency_comparsion_L10_10}, respectively. The molecular simulations are performed without applying PBCs. As Fig.~\ref{fig:Gra_frequency_comparsion_L10_10} shows, the frequency vs. stretch curves obtained from the two approaches agree within $ \approx 95\%$. It is observed that for all the three loading cases, the frequencies increase monotonically with the stretch. The rate of increase in the frequency is higher at small stretches than at higher stretches. \citet{Singh2018} have reported a similar behaviour for a rectangular SLGS under uniaxial stretch. It is also observed that at zero strain, the frequencies from the molecular simulations are $\approx 15\%$ higher than those computed from the continuum model. This difference may be due to residual stresses in the relaxed configuration of SLGS, which are not accounted for in the present continuum model. For uniaxial stretch along the armchair direction, the frequencies from the molecular simulations and those from the continuum model agree up to $\lambda_1= 1.15$ after which the MS simulation results show a stiffer behavior. This is consistent with the variation in the stress with stretch shown in Figs.~\ref{fig:long_stress_AC} and \ref{fig:long_stress_ZZ}. At higher stretches, the variation in the frequency is higher when the SLGS is stretched along the armchair direction than in the zigzag direction. This mild anisotropic response is attributed to the angle-bending energy, which contributes significantly to the total potential energy when the SLGS is stretched along the armchair direction. Contrary to the molecular simulation results, the continuum model shows the zigzag direction to be stiffer\footnote{I.e.~the mode shapes with a higher number of sinusoidal waves along the stiffer direction have a higher frequencies than the one with higher waves along the softer direction.}.
\subsubsection{Carbon nanocone}
The modal analysis of a pre-stretched simply-supported CNC is carried out using MS simulations employing the MM3 potential. The modal frequencies obtained from the MS simulations are compared with those obtained from the continuum model at each stretch. For this study, a truncated cone is selected with initial apex angle $\alpha = 14.2^\circ$, height $H = 12.4$ nm, and tip radius $R = 1.017$ nm.
\begin{figure}[!htbp]
    \begin{subfigure}[t]{0.49\linewidth}
        \centering
 \includegraphics[height=65mm]{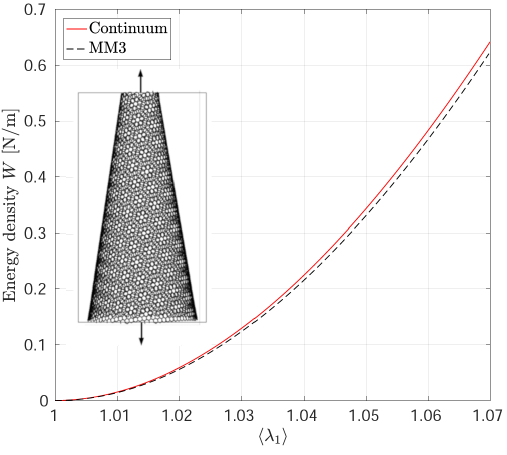}
         \vspace{-3mm}
        \subcaption{}
    \end{subfigure}
    \begin{subfigure}[t]{0.49\linewidth}
        \centering
    \includegraphics[height=65mm]{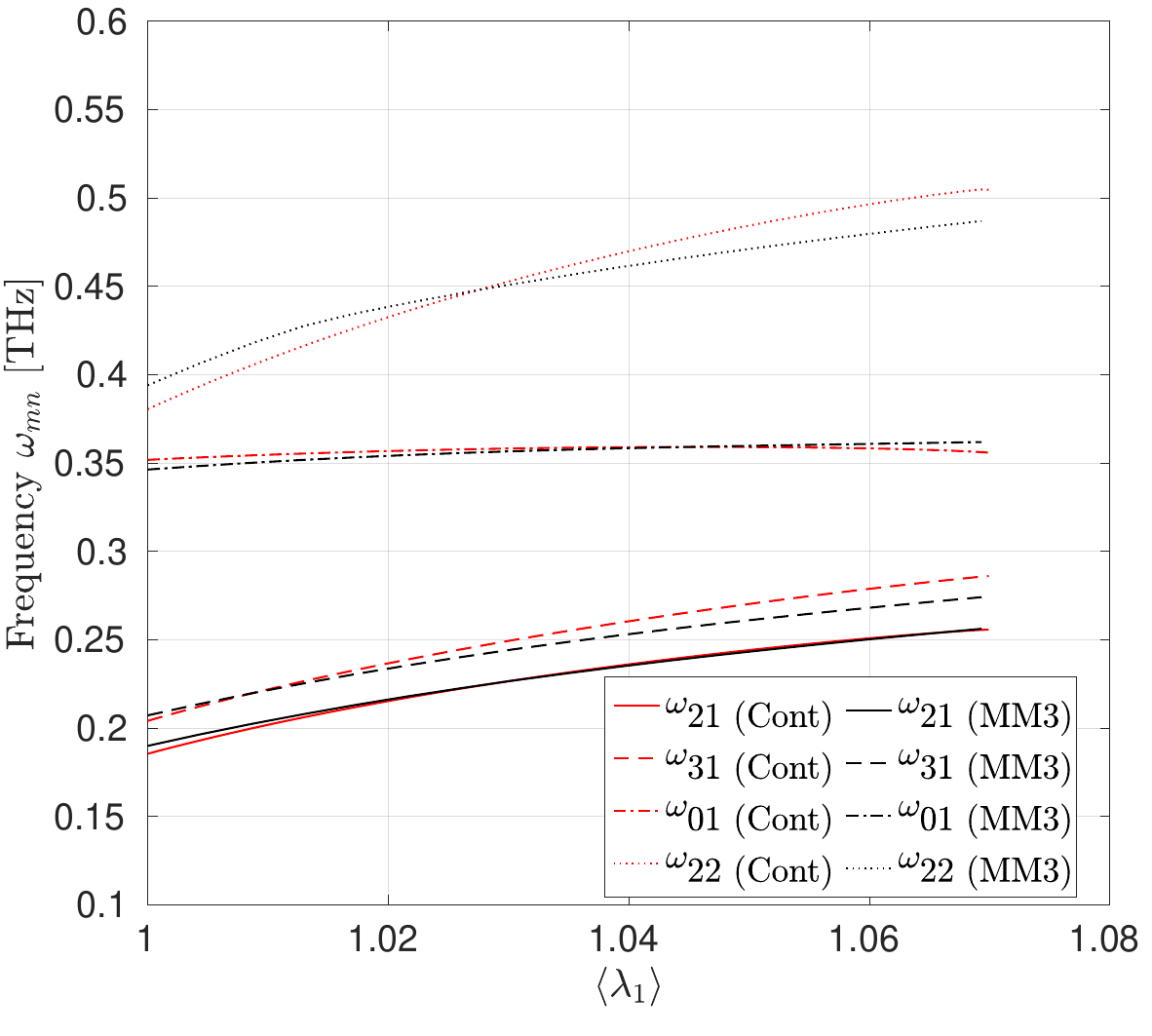}
        \vspace{-3mm}
        \subcaption{}
    \end{subfigure}
     \caption{Variation in the total potential energy (a) and frequencies (b) of a CNC under uniaxial stretch. Here, $\langle\lambda_1\rangle$ denotes the average stretch along the CNC axis. \label{fig:cnc}}
\end{figure}

\begin{figure}[!htbp]
        \centering
    \begin{subfigure}[t]{0.2\linewidth}
        \centering
    \includegraphics[height=65mm]{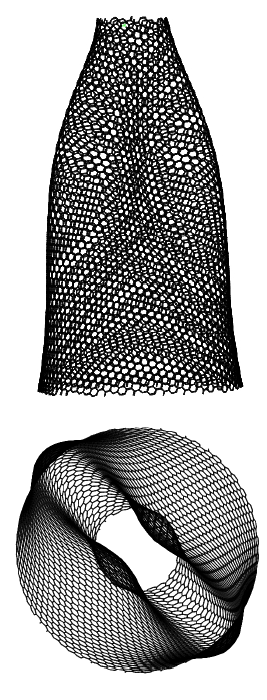}
        \vspace{-3mm}
        \subcaption{}
    \end{subfigure}
    \begin{subfigure}[t]{0.2\linewidth}
        \centering
    \includegraphics[height=65mm]{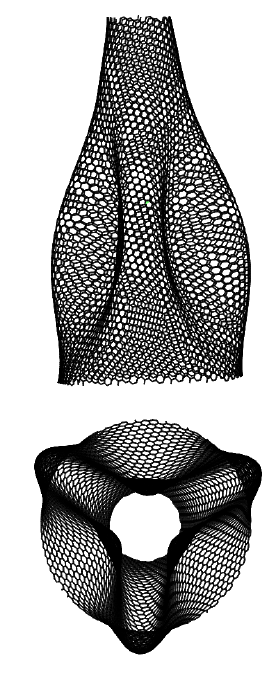}
        \vspace{-3mm}
        \subcaption{}
    \end{subfigure}
    \begin{subfigure}[t]{0.2\linewidth}
        \centering
     \includegraphics[height=65mm]{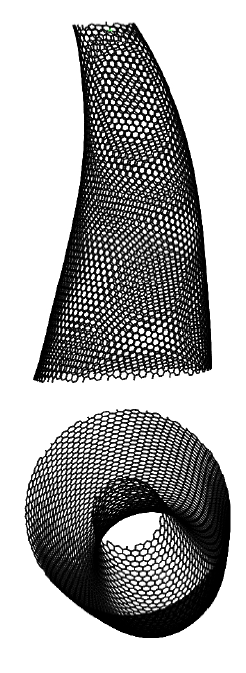}
         \vspace{-3mm}
        \subcaption{}
    \end{subfigure}
    \begin{subfigure}[t]{0.2\linewidth}
        \centering
     \includegraphics[height=65mm]{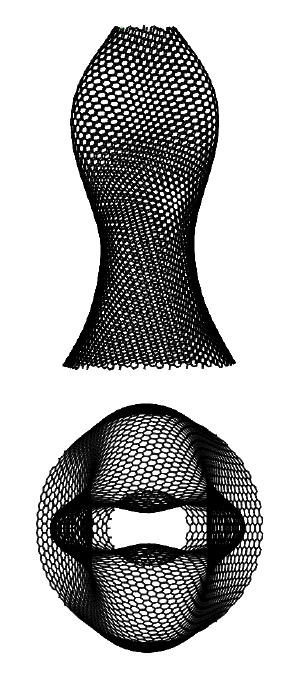}
         \vspace{-3mm}
        \subcaption{}
    \end{subfigure}\\
    \vspace{-3mm}
    \caption{Mode shapes of a relaxed simply-supported carbon nanocone determined by MS simulations: Shape corresponding to frequency (a) $\omega_{21}$, (b) $\omega_{31}$, (c) $\omega_{01}$ and (d) $\omega_{22}$. Here the two indices in $\omega$ represent the number of circumferential waves and axial half waves in that order.\label{fig:cnc_modes}}
\end{figure}

Before computing the modal frequencies in the MS simulations, the minimum energy configuration of the structure is found at each stretch level. The stretch is obtained by fixing all the degrees-of-freedom of the edge atoms of the larger radius and incrementally moving the atoms of the smaller radius by 0.1 \AA \, in the axial direction. We note that this leads to an inhomogeneous stretch state in the CNC, due to its tapered geometry. The following results are therefore taken at the average stretch $\langle\lambda_1\rangle = h/H$, where $h$ is the current height. The variations in the potential energy and the modal frequencies computed from the MM3 potential and the continuum model are shown in Fig.~\ref{fig:cnc}. The corresponding mode shapes computed from the MM3 potential are shown in Fig.~\ref{fig:cnc_modes}.
For moderate values of the stretch, the potential energy from both the approaches are in good agreement. As in the case of SLGS, the modal frequencies from both the approaches differ slightly in the unstretched configuration, possibly due to the absence of residual stresses in the continuum model. The frequency of the lower order modes $\omega_{01}$ and $\omega_{21}$ obtained from the MM3 potential and the continuum model are in good agreement. Based on continuum shell theory, we conjecture that this agreement is due to the dominant membrane/stretching deformation in the lower modes for which the present continuum model is calibrated. This is also the reason for a good match in the strain energy since for moderate value of $\langle\lambda_{1}\rangle$ only the membrane deformations are dominating. At higher modes bending deformation dominates and hence there is a disagreement between the frequencies computed from the two approaches. Finally, we note that $\omega_{01}$ remains unaffected for the range of $\langle\lambda_{1}\rangle$ considered here.
\subsection{Summary of main findings} \label{main_f}
While comparing the elastic response of SLGS obtained from the three potentials and the continuum model with that from DFT, we find the following: \\
1) The MM3 potential results agree up to a stretch of 1.1 (10$\% $ strain).\\
2) The REBO+LJ potential results agree only within the range of small deformation.\\
3) The two different parameterizations of the Tersoff potential fail to capture the behavior correctly. Further, they predict negative Poisson ratio. \\
4) The agreement of the energies and stresses for uniaxial and biaxial stretch between the continuum model and the DFT model are within an error of $ \approx 0.05 \%$. \\
While examining the vibrations of stretched SLGS and CNC studied using MM3 potential and DFT based continuum model, we find that: \\
5) The transverse vibrational frequencies increase monotonically with the pre-stretch. \\
6) The frequencies found in MS simulations differ from those obtained using the continuum model at zero stretch.\\
7) The bending frequency $\omega_{01}$ of CNC is nearly unaffected by the range of stretch studied in the paper.
\section{Conclusions} \label{Conclusions}
The nonlinear mechanical response of square SLGS under large uniaxial stretch and pure dilatation is investigated with continuum and atomistic simulations employing the MM3, REBO+LJ, and Tersoff potentials. The obtained results are compared with DFT results available in the literature for the two stretch states. The elastic response obtained from the continuum model is in good agreement with DFT results. Among the potentials used for molecular simulations, the MM3 is found to be the most accurate in this study. The transverse vibrational frequencies of a square SLGS and a CNC found from the continuum model and molecular simulations employing the MM3 potential are found to be in good agreement, within an error of about $~5\% $. In future work, the present study will be extended to finite temperatures.\\
\section*{Acknowledgement}{Financial support from the German Research Foundation (DFG)
through grants SA 1822/8-1 and GSC 111 is gratefully acknowledged. The authors thank Farzad Shirazian for his comments on the atomistic simulations.}
\appendix
\section{Description of the molecular potentials}
In this section, the mathematical expressions of the used potentials are given.
\subsection{MM3 Potential} \label{mm3-1}
The terms in Eq.~(\ref{eq:mm3}) are given for atoms $I$, $J$, $K$ and $L$ by
\eqb{lll}
\notag
U_{\text{s}}\is \ds 71.94 K_{\text{s}} (r_{IJ}-{r^{\text{e}}})^2\left[1-2.55(r_{IJ}-r^{\text{e}})+\left(\frac{7}{12}\right)2.55(r_{IJ}-r^{\text{e}})^2 \right]~,\\[3mm]
U_{\theta} \is \ds 0.021914K_\theta (\theta_{IJK} - {\theta^{\text{e}}})^2 \biggl[1-0.014(\theta_{IJK} - {\theta^{\text{e}}})+5.6(10)^{-5}(\theta_{IJK} - {\theta^{\text{e}}})^2 \\[2mm]
\mi \ds 7.0(10)^{-7}(\theta_{IJK}-{\theta^{\text{e}}})^3+9.0(10)^{-10}(\theta_{IJK} - {\theta^{\text{e}}})^4 \biggr]~,\\[3mm]
U_{\phi}\is\ds \frac{V_1}{2}(1+\cos \phi_{IJKL} )+\frac{V_2}{2}(1-\cos 2\phi_{IJKL}) +\frac{V_3}{2}(1+\cos 3\phi_{IJKL}) ~,\\[3mm]
U_{\text{s}\theta}\is\ds 2.51118K_{\text{s}\theta} \left[ (r_{IJ}-{r^{\text{e}}})+(r_{JK}-{r^{\text{e}}}) \right](\theta_{IJK} -{\theta^{\text{e}}})~,\\[3mm]
\eqe
\eqb{lll}
U_{\phi \text{s}}\is\ds 11.995 \frac{K_{\phi \text{s}}}{2} (r_{IJ}-{r^{\text{e}}})(1+\cos (3 \phi_{IJKL}))~,\\[3mm]
U_{\theta \theta^{'}}\is\ds -0.021914K_{\theta \theta^{'}}(\theta_{IJK} - {\theta^{\text{e}}})(\theta_{IKL} -{\theta^{\text{e}}})~,\\[3mm]
U_{\text{vdW}}\is\ds \epsilon_{\text{e}}\left [-2.25\left(\frac{r_{\text{v}}}{r_{IJ}}\right)^6+1.84(10)^5\text{exp}\left(-12.0\frac{r_{IJ}}{r_{\text{v}}}\right) \right] ~,
\eqe
where $r_{IJ}$, $\theta_{IJK}$ and $\phi_{IJKL}$ are the bond length, the angle between the bonds and the torsion angle, respectively. The parameters with superscript e define the equilibrium values at which the total potential energy is minimum. $K_{\text{s}}$, $K_\theta$, $V_1$, $V_2$, $V_3$, $\epsilon_{\text{e}}$, $r_{\text{v}}$, $K_{\text{s} \theta}$, $K_{\phi s}$, $K_{\theta \theta^{'}}$, {$r^{\text{e}}$} and  {$\theta^{\text{e}}$} are the potential parameters and are listed in Table~\ref{tab:mm3}.
\begin{table}[h]
\footnotesize
\centering
\caption{The potential parameters of MM3 \citep{Allinger1989}. \label{tab:mm3}}
\begin{tabular}{lcccc}
  \hline
    Parameter & value  \\
    \hline
   $K_{\text{s}}$   & 4.49 mdyne/\AA \\
   $K_\theta$   & 0.67 mdyne-\AA/rad$^2$  \\
   $V_1$ & 0.185 Kcal/mol  \\
   $V_2$ & 0.170 Kcal/mol  \\
   $V_3$ & 0.520 Kcal/mol  \\
   $\epsilon_{\text{e}}$ & 0.027 Kcal/mol  \\
   $r_{\text{v}}$ & 2.04 \AA  \\
   $K_{\text{s} \theta}$ & 0.130 mdyne/rad  \\
   $K_{\phi s}$ & 0.059 mdyne/rad  \\
   $K_{\theta \theta^{'}}$ & 0.240 mdyne-\AA/rad$^2$  \\
   ${r^{\text{e}}}$ & {1.5247 \AA} \\
   ${\theta^{\text{e}}} $ & {110.2\degree} \\
  \hline
\end{tabular}
\end{table}
\subsection{REBO+LJ Potential} \label{reb-1}
The repulsive ($E_{\text{R}}$) and attractive ($E_{\text{A}}$) terms in Eq.~(\ref{eq:rebo}) are
\eqb{lll}
 E_{\text{R}}(r_{IJ})=\ds {w(r_{IJ})}\left(1+\frac{Q}{r_{IJ}}\right)\,A\,e^{-\alpha \, r_{IJ}}~,
\eqe
\eqb{lll}
 E_{\text{A}}(r_{IJ}) =\ds -{w(r_{IJ})} \sum_{n=1}^{3}\,B^n\,e^{-\beta^n \, r_{IJ}}~,
\eqe
where $Q$, $A$, $\alpha$, $B^{n}$ and $\beta^{n}$ are the potential parameters that depend on the type of atoms $I$ and $J$ and are given in Table~\ref{tab:rebop}. {$w$} is the bond-weighting factor that depends on the cutoff distances (${r^{\mathrm{min}}}$ and ${r^{\mathrm{max}}}$) and varies from zero to one.  The REBO interactions are gradually turned off, when the bond length is in the range ${r^{\mathrm{min}}} < r_{IJ} < {r^{\mathrm{max}}}$. This is achieved through the bond-weighting factor
\eqb{lll} \label{eq:re_cut}
\ds {w(r_{IJ})} \is \ds S^{'}(t_{\text{c}}(r_{IJ}))~,
\eqe
where the switching and scaling functions $S^{'}$ and $t_c$
\eqb{lll}
\ds S'(t) \is \ds \Theta(-t)+\frac{1}{2}\Theta(-t)\Theta(1-t)\left(1+\cos(\pi t)\right)~,
\eqe
\eqb{lll}
\ds t_{\text{c}}(r_{IJ}) \is \ds \frac{r_{IJ}-{r^{\text{min}}}}{{r^{\text{max}}}-{r^{\text{min}}}}~.
\eqe
Here $\Theta(t)$ is the Heaviside step function.

\begin{table}[h]
\footnotesize
\centering
\caption{The REBO+LJ potential parameters for carbon \citep{Stuart2000}. \label{tab:rebop}}
\begin{tabular}{lcccc}
  \hline
    Parameter & value  \\
    \hline
   $Q$ (\AA)   & 0.313460 \\
   $\alpha$ (\AA $^{-1}$)  & 4.746539  \\
   $A$ (eV) & 10953.544 \\
   $B^1$ (eV) & 12388.792  \\
   $B^2$ (eV) & 17.567065  \\
   $B^3$ (eV) & 30.714932  \\
   $\beta^1$ (\AA $^{-1}$) & 4.720452  \\
   $\beta^2$ (\AA $^{-1}$) & 1.433213  \\
   $\beta^3$ (\AA $^{-1}$) & 1.382691  \\
   $\epsilon$ (eV) & 0.002840  \\
   $\sigma$ (\AA) & 3.4     \\
   ${r^{\text{min}}}$ (\AA) & 1.7  \\
   ${r^{\text{max}}}$ (\AA) & 2.0  \\
  \hline
\end{tabular}
\end{table}
\subsection{Tersoff Potential} \label{ter-1}
The terms in the Tersoff potential of Eq.~(\ref{eq:tersoff}) are given by
\eqb{lll}
 f_{\text{R}}(r_{IJ})=\ds A \, e^{-\lambda \, r_{IJ}}~,
\eqe
\eqb{lll}
 f_{\text{A}}(r_{IJ}) =\ds -B \, e^{-\mu \, r_{IJ}}~,
 \eqe
and
\eqb{lll}
f_{\text{c}}(r):=\left\{
\begin{matrix}
   1     & r_{IJ}<{r^{\mathrm{min}}}~,  \\
   \ds \frac{1}{2} \left[1+\cos \left(\frac{(r_{IJ}-{r^{\mathrm{min}}})\pi}{{r^{\mathrm{max}}}-{r^{\mathrm{min}}}}\right)\right]     & {r^{\mathrm{min}}}<r_{IJ}<{r^{\mathrm{max}}}~,  \\
    0       & r_{IJ}>{r^{\mathrm{max}}}~.
\end{matrix} \right. \label{tersoff_cut}
\eqe
where $A$, $B$, $\lambda$ and $\mu$ are the parameters that depend on paired atom types. 
\begin{table}[h]
\footnotesize
\centering
\caption{The Tersoff and modified Tersoff potential parameters for carbon. \label{tab:tersoff}}
\begin{tabular}{lcccc}
  \hline
     Parameter & \citet{Tersoff1989} & \citet{Albe2005}\\
    \hline
   $A$ (eV)  & $1393.600  $ & 2019.800 \\
   $B$  (eV) & $346.700$  & 175.400 \\
   $\lambda$ (\AA $^{-1}$) & 3.487 & 4.184 \\
   $\mu$ (\AA $^{-1}$) & 2.2119 & 1.931 \\
   $\beta$ & $1.5724 \times 10^{-7}$ & 1.000 \\
   $n$ & 0.727  & 1.000\\
   $c$ & 38049.00 & 181.910 \\
   $d$ & $4.384 $ & 6.284\\
   $h$ & -0.57058 & -0.5556 \\
   ${r^{\text{min}}}$ (\AA ) & 1.80 & 1.85 \\
   ${r^{\text{max}}}$ (\AA ) & 2.10 & 2.15 \\
   $\chi$ & 1.0 & 1.0\\
   $\omega$ & 1.0 & 1.0\\
  \hline
\end{tabular}
\end{table}
The cutoff function ($f_{\text{c}}$) describes a gradual decrease of the bond strength between ${r^{\mathrm{min}}}$ and ${r^{\mathrm{max}}}$. In Eq. (\ref{eq:tersoff}), $b_{IJ}$ is the bond order term. It depends on neighboring atoms and it is defined by
\eqb{lll}
 b_{IJ} \dis \ds \chi\left(1+\beta^{n}\zeta_{IJ}^{n}\right)^{\frac{-1}{2\,n}}~,
\eqe
\eqb{lll}
 \zeta_{IJ}=\ds \sum_{K\neq I,J}f_\text{c}(r_{IK})\,\omega\,g(\theta_{IJK})~,
\eqe
\eqb{lll}
g(\theta_{IJK})=\ds 1+\frac{c^2}{d^2}-\frac{c^2}{d^2+(h-\cos(\theta_{IJK}))^2}~,
\eqe
where $\theta_{IJK}$ is the angle between bond $IJ$ and $IK$, while $A$, $B$, $\lambda$, $\mu$, $\beta^{n}$, $n$, $c$, $d$, $h$, ${r^{\text{min}}}$, ${r^{\text{max}}}$, $\chi$ and $\omega$ are the potential parameters that for carbon are given in Table~\ref{tab:tersoff}.
\bibliographystyle{unsrtnat}
\bibliography{Bib/bibliographym}

\end{document}